\documentclass[preprint]{ptephy}

\preprintnumber{KYUSHU-HET-196}

\usepackage{amssymb}
\usepackage{amsthm}
\usepackage{amsmath}
\usepackage{booktabs}
\usepackage{bbm}
\usepackage{mathtools}
\DeclareMathOperator{\tr}{tr}
\DeclareMathOperator{\Tr}{Tr}

\DeclareMathOperator{\Ln}{Ln}

\newcommand{\Slash}[1]{{\ooalign{\hfil/\hfil\crcr$#1$}}}
\numberwithin{equation}{section}

\begin{document}

\title{Infrared renormalon in the supersymmetric $\mathbb{C}P^{N-1}$ model on $\mathbb{R}\times S^1$}

\author{%
\name{\fname{Kosuke} \surname{Ishikawa}}{1},
\name{\fname{Okuto} \surname{Morikawa}}{1},
\name{\fname{Akira} \surname{Nakayama}}{1},
\name{\fname{Kazuya} \surname{Shibata}}{1},\\
\name{\fname{Hiroshi} \surname{Suzuki}}{1,\ast}, and
\name{\fname{Hiromasa} \surname{Takaura}}{1}
}

\address{%
\affil{1}{Department of Physics, Kyushu University
744 Motooka, Nishi-ku, Fukuoka, 819-0395, Japan}
\email{hsuzuki@phys.kyushu-u.ac.jp}
}

\date{\today}

\begin{abstract}
In the leading order of the large-$N$ approximation, we study the renormalon
ambiguity in the gluon (or, more appropriately, photon) condensate in the 2D
supersymmetric $\mathbb{C}P^{N-1}$ model on~$\mathbb{R}\times S^1$ with the
$\mathbb{Z}_N$ twisted boundary conditions. In our large~$N$ limit, the
combination $\Lambda R$, where $\Lambda$ is the dynamical scale and $R$~is the
$S^1$ radius, is kept fixed (we set $\Lambda R\ll1$ so that the perturbative
expansion with respect to the coupling constant at the mass scale~$1/R$ is
meaningful). We extract the perturbative part from the large-$N$ expression of
the gluon condensate and obtain the corresponding Borel transform~$B(u)$.
For~$\mathbb{R}\times S^1$, we find that the Borel singularity at~$u=2$, which
exists in the system on the uncompactified~$\mathbb{R}^2$ and corresponds to
twice the minimal bion action, disappears. Instead, an unfamiliar renormalon
singularity \emph{emerges\/} at~$u=3/2$ for the compactified
space~$\mathbb{R}\times S^1$. The semi-classical interpretation of this
peculiar singularity is not clear because $u=3/2$ is not dividable by the
minimal bion action. It appears that our observation for the system
on~$\mathbb{R}\times S^1$ prompts reconsideration on the semi-classical bion
picture of the infrared renormalon.
\end{abstract}

\subjectindex{B06, B16, B32, B34, B35}
\maketitle

\section{Introduction}
\label{sec:1}
In~Refs.~\cite{Argyres:2012vv,Argyres:2012ka,Dunne:2012ae,Dunne:2012zk}, it was
claimed that the ambiguity in perturbation theory caused by the infrared (IR)
renormalon~\cite{tHooft:1977xjm,Beneke:1998ui}---a single Feynman diagram whose
amplitude grows factorially as a function of the order---is canceled by the
ambiguity associated with a semi-classical object called the (neutral) bion---a
pair of the fractional instanton~\cite{Eto:2004rz,Eto:2006mz,Eto:2006pg,%
Bruckmann:2007zh,Brendel:2009mp,Bruckmann:2018rra} and the fractional
anti-instanton. This cancellation mechanism between perturbation theory and a
semi-classical object is analogous to the cancellation between the ambiguity
caused by the proliferation of the number of Feynman
diagrams~\cite{Brezin:1976wa,Lipatov:1976ny} (see also
Ref.~\cite{LeGuillou:1990nq} for a review) and the ambiguity associated with
the instanton--anti-instanton pair~\cite{Bogomolny:1980ur,ZinnJustin:1981dx}.
Such a cancellation of the ambiguity will be crucial for the enterprise of a
fully semi-classical understanding of the low-energy physics of gauge theory;
see~Ref.~\cite{Dunne:2015eaa} and references therein. A crucial element in such
a semi-classical argument is an $S^1$~compactification; the $S^1$ radius~$R$
provides a mass scale and if it is sufficiently small compared to the dynamical
scale~$\Lambda$, $\Lambda R\ll1$, it would allow a semi-classical (weak
coupling) treatment. On the other hand, a non-trivial holonomy along~$S^1$, or,
equivalently, twisted boundary conditions along~$S^1$, allow the fractional
(anti-)instanton.

For the 4D $SU(N)$ gauge theory, the above cancellation between the IR
renormalon and the bion is still conjectural. In particular, the ambiguity
associated with the bion does not exactly correspond to that of the renormalon
on~$\mathbbm{R}^4$, which is typically characterized by~$e^{-2S_I/(N\beta_0)}$,
where $S_I$ is the instanton action and $\beta_0=11/3-2n_W/3$~is the
coefficient of the one-loop beta function for the 't~Hooft coupling; $n_W$~is
the number of Weyl fermions in the adjoint representation. It remains a
non-trivial task to investigate if the ambiguity due to the bion coincides with
the above form by possible renormalization and de-compactification effects;
see also Ref.~\cite{Hongo:2018rpy}. For the 2D $\mathbb{C}P^{N-1}$ model, which
shares many similarities with 4D gauge theory, the beta function is
simply~$\beta_0=1$ and one might further push the above semi-classical
interpretation of the renormalon.

In a recent interesting paper~\cite{Fujimori:2018kqp}, the authors carried out
a systematic investigation on this issue in the 2D supersymmetric
$\mathbb{C}P^{N-1}$ model on~$\mathbb{R}\times S^1$ with the $\mathbb{Z}_N$
twisted boundary conditions. They computed a one-loop-order effective action
for quasi-collective coordinates associated with the bion configuration. Then,
by employing the Lefschetz thimble
method~\cite{Witten:2010cx,Cristoforetti:2012su}, they computed the bion
contribution to the vacuum energy as a function of the supersymmetry breaking
parameter~$\delta\epsilon$. They found that the bion induces ambiguity in the
$O(\delta\epsilon^2)$ term of the vacuum energy with the
strength~$\sim e^{-2S_I/N\beta_0}$; they then inferred that this ambiguity is
canceled by the ambiguity caused by the IR renormalon.

Although the above computation is very explicit, the interpretation of the
result may be disputable. It is generally not well understood whether the
ambiguity due to the bion truly corresponds to the renormalon ambiguity. In
particular, there is a study~\cite{Anber:2014sda} that sounds incompatible with
the above result; this study asserts that in gauge theory
on~$\mathbb{R}^3\times S^1$, the $S^1$ compactification works as an IR cutoff
and, as a consequence, the IR renormalon disappears in an $S^1$ compactified
space. If we assume this mechanism to be general, what cancels the ambiguity
caused by the bion? On the other hand, there is also a study where the
ambiguity due to the bion cancels the perturbative ambiguity.
In~Refs.~\cite{Fujimori:2016ljw,Fujimori:2017oab}, it was found that the bion
and perturbative ambiguities in the ground-state energy indeed have the same
magnitude with opposite signs in an (approximately) supersymmetric
$\mathbb{C}P^1$ quantum mechanics; this system is obtained by a reduction of
the 2D $\mathcal{N}=(2,0)$ supersymmetric $\mathbb{C}P^1$ model to lower-lying
Kaluza--Klein (KK) momentum modes.

To have some insight into this confused situation and investigate the relation
between the bion and the renormalon more directly, it would be useful to
consider the large-$N$ limit (see Ref.~\cite{Coleman:1985rnk} for a classical
exposition). The large-$N$ limit distinguishes the instanton and the renormalon
because their effects differ by the factor~$N$. Typically, as the matrix model
illustrates~\cite{Brezin:1977sv}, the perturbative series tends to become a
convergent series in the large-$N$ limit, possibly leaving the effect of
renormalons. Motivated by this reasoning, in this paper, we consider the large
$N$ approximation of the 2D supersymmetric $\mathbb{C}P^{N-1}$ model
on~$\mathbb{R}\times S^1$ with the $\mathbb{Z}_N$ twisted boundary conditions
and investigate the IR renormalon via systematic calculations. For the system
in the uncompactified space $\mathbb{R}^2$, the large-$N$ solution is
well known~\cite{DAdda:1978dle}. We generalize this solution to the
compactified space, $\mathbb{R}\times S^1$.

In our large-$N$ limit, we assume
\begin{equation}
   \text{$\Lambda R=\text{const.}$ as $N\to\infty$},
\label{eq:(1.1)}
\end{equation}
where $\Lambda$ is a dynamical scale (i.e., the $\Lambda$~parameter) and $R$ is
the $S^1$ radius.\footnote{We assume this for technical reasons. An alternative
large-$N$ limit, in which
\begin{equation}
   \text{$\Lambda RN=\text{const.}$ as $N\to\infty$},
\label{eq:(1.2)}
\end{equation}
might be natural for a semi-classical consideration, because the potential
height between $N$~degenerate classical vacua of the present system is
characterized by the scale~$1/(RN)$. We could not find, however, a convincing
way to analyze the resulting expressions in this large-$N$ limit.} We also
assume that $\Lambda R$ is sufficiently small ($\Lambda R\ll1$) so that the
perturbative expansion with respect to the coupling constant at the mass
scale~$1/R$ is meaningful.

In this paper, as a simple and non-trivial quantity, we study the gluon (or
more appropriately, photon) condensate in the above system in the leading order
of the large-$N$ approximation. We use the following definitions in studying a
factorially divergent series. For the perturbative series of a
quantity~$f(\lambda)$,
\begin{equation}
   f(\lambda)\sim\sum_{k=0}^\infty f_k\left(\frac{\lambda}{4\pi}\right)^{k+1},
\label{eq:(1.3)}
\end{equation}
where $\lambda$ denotes the 't~Hooft coupling constant, we define the Borel
transform by
\begin{equation}
   B(u)\equiv\sum_{k=0}^\infty\frac{f_k}{k!}u^k.
\label{eq:(1.4)}
\end{equation}
Then the Borel sum is given by
\begin{equation}
   f(\lambda)\equiv\int_0^\infty du\,B(u)\,e^{-4\pi u/\lambda}.
\label{eq:(1.5)}
\end{equation}
If the perturbative coefficient~$f_k$ in~Eq.~\eqref{eq:(1.3)} grows
factorially~$f_k\sim b^{-k}k!$ as~$k\to\infty$, the Borel
transform~$B(u)$~\eqref{eq:(1.4)} develops a singularity located at~$u=b$. If
this singularity is on the positive real $u$ axis, $b>0$, then the Borel
integral~\eqref{eq:(1.5)} becomes ill defined and produces an ambiguity
proportional to~$\sim e^{-4\pi b/\lambda}$. In this convention, the IR renormalon
in the present system in the large-$N$ limit is generally expected to produce
Borel singularities at positive integers $u=1$, $2$, \dots. Since the minimum
action of the bion is~$4\pi/\lambda$ (when the constituent fractional instanton
and the anti-instanton are infinitely separated), the ambiguity caused by the
bion for the vacuum energy~\cite{Fujimori:2018kqp} corresponds to~$u=1$. On the
other hand, as we will see below, the gluon condensate in the system
in~$\mathbb{R}^2$ suffers from the IR renormalon at~$u=2$; the associated
factor in the exponential, $8\pi/\lambda$, is twice the minimum bion action.

In this paper, we show that for the gluon condensate in the compactified
space~$\mathbb{R}\times S^1$, the renormalon singularity at~$u=2$, which exists
for~$\mathbb{R}^2$, disappears. This sounds consistent with the claim
in~Ref.~\cite{Anber:2014sda} that there is no IR renormalon in an $S^1$
compactified space. However, for~$\mathbb{R}\times S^1$, we find that an
unfamiliar renormalon singularity \emph{emerges\/} at~$u=3/2$. Our observation
thus indicates that an $S^1$ compactification significantly affects the
renormalon structure. Furthermore, we do not know any semi-classical
interpretation of this singularity. It appears that our finding prompts
reconsideration on the above semi-classical picture of the IR renormalon.

This paper is organized as follows: In~Sect.~\ref{sec:2}, we define our system
starting from the expressions in~Ref.~\cite{Fujimori:2018kqp} but using the
't~Hooft coupling~$\lambda\equiv g^2N$ (the 't~Hooft coupling is extensively
used in this paper). For the large-$N$ approximation, it is highly convenient
to employ the homogeneous coordinate of~$\mathbb{C}P^{N-1}$ and its
``superpartner''; these are introduced in~Sect.~\ref{sec:2.2}. Then, as a
standard procedure in the large-$N$ approximation, we introduce auxiliary
fields to impose constraints among original $N$~fields and to make the action
quadratic in the $N$~fields. In~Sect.~\ref{sec:3}, we obtain the effective
action for the auxiliary fields by integrating over the $N$~fields. We find the
saddle point in the large~$N$ limit~\eqref{eq:(1.1)} and then compute the
effective action for fluctuations around the saddle point to the quadratic
order. In~Sect.~\ref{sec:4}, we compute the gluon condensate in the leading
order of the large~$N$ approximation. We then extract the perturbative part
from the large-$N$ expression and obtain the corresponding Borel transform. By
studying singularities of the Borel transform, we arrive at the above
conclusion. We also argue that a perturbative calculation of a physical
observable inherits the renormalon ambiguity found in the gluon condensate.
Section~\ref{sec:5} is devoted to the conclusion and discussion.
Appendix~\ref{sec:A} contains some useful formulas to translate expressions in
the inhomogeneous coordinate to those in the homogeneous coordinate.
Appendix~\ref{sec:B} gives some rigorous bounds on functions appearing in the
effective action. In Appendix~\ref{sec:C}, we explain in some detail how
the chiral symmetry breaking term~$S_{\text{local}}$~\eqref{eq:(3.39)} emerges
with our convention for~$\gamma_5$ in the dimensional regularization.

\section{$\mathcal{N}=(2,2)$ supersymmetric $\mathbb{C}P^{N-1}$ model on
$\mathbb{R}\times S^1$}
\label{sec:2}
\subsection{Definition of the system}
We suppose that the spacetime is~$\mathbb{R}\times S^1$ and denote the
coordinate of~$\mathbb{R}$ by~$x$ and that of~$S^1$ by~$y$; the radius of~$S^1$
is~$R$ and thus $0\leq y<2\pi R$. The Greek indices $\mu$, $\nu$, etc.\ run
over $x$ and~$y$. We start with~Eq.~(4.1) of~Ref.~\cite{Fujimori:2018kqp} in a
somewhat different notation:\footnote{Unless stated otherwise, the summation
over repeated indices is always understood. This $\mathcal{N}=(2,2)$ action can
be obtained by the dimensional reduction of the 4D $\mathcal{N}=1$ Wess--Zumino
model~\cite{Wess:1992cp} by setting
$\Phi^a\equiv\varphi^a+\sqrt{2}\theta\psi+\theta\theta F$, where
$\psi_1^a=(\psi_+^a+\psi_-^a)/\sqrt{2}$,
$\psi_2^a=(-\psi_+^a+\psi_-^a)/\sqrt{2}$,
$\Bar{\psi}_{\Dot{1}}^a=(\Bar{\psi}_+^a+\Bar{\psi}_-^a)/\sqrt{2}$, and
$\Bar{\psi}_{\Dot{2}}^a=(-\Bar{\psi}_+^a+\Bar{\psi}_-^a)/\sqrt{2}$.
The K\"ahler potential is taken as
$K=\frac{N}{\lambda}\ln(1+\sum_{a=1}^{N-1}\Bar{\Phi}^a\Phi^a)$. Note that we are
working in the Euclidean space (the Lorentzian time is given by~$x^0=-ix$); the
Boltzmann weight in the functional integral is thus~$e^{-S}$.}
\begin{align}
   S&\equiv\int d^2x\,\frac{2N}{\lambda}
   \biggl\{
   G_{a\Bar{b}}
   \Bigl[
   \partial\varphi^a\Bar{\partial}\Bar{\varphi}^{\Bar{b}}
   +\Bar{\partial}\varphi^a\partial\Bar{\varphi}^{\Bar{b}}
\notag\\
   &\qquad\qquad\qquad\qquad\qquad{}
   +\Bar{\psi}_-^{\Bar{b}}
   (\partial\psi_-^a+{\mit\Gamma}^a{}_{cd}\partial\varphi^c\psi_-^d)
   +\psi_+^a
   (\Bar{\partial}\Bar{\psi}_+^{\Bar{b}}
   +{\mit\Gamma}^{\Bar{b}}{}_{\Bar{c}\Bar{d}}
   \Bar{\partial}\Bar{\varphi}^{\Bar{c}}\Bar{\psi}_+^{\Bar{d}})
   \Bigr]
\notag\\
   &\qquad\qquad\qquad\qquad\qquad{}
   -\frac{1}{2}R_{a\Bar{b}c\Bar{d}}
   \psi_+^a\Bar{\psi}_+^{\Bar{b}}\psi_-^c\Bar{\psi}_-^{\Bar{d}}
   \biggr\}
\notag\\
   &\qquad{}+S_{\text{top}}.
\label{eq:(2.1)}
\end{align}
In this paper, we use the 't~Hooft coupling~$\lambda$ that is defined
by~$\lambda\equiv g^2N$ from the coupling constant~$g$
in~Ref.~\cite{Fujimori:2018kqp}. The lowercase indices $a$, $\Bar{b}$ etc.\ run
over $1$, \dots, $N-1$, and
\begin{align}
   G_{a\Bar{b}}
   &\equiv
   \frac{\partial^2}{\partial\varphi^a\partial\Bar{\varphi}^{\Bar{b}}}
   \ln\left(1+\sum_{c=1}^{N-1}|\varphi^c|^2\right)
\notag\\
   &=\frac{\delta^{ab}}{1+\sum_c|\varphi^c|^2}
   -\frac{\Bar{\varphi}^{\Bar{a}}\varphi^b}
   {\left(1+\sum_c|\varphi^c|^2\right)^2}
\label{eq:(2.2)}
\end{align}
is the Fubini--Study metric on~$\mathbb{C}P^{N-1}$. The connection and the
curvature on~$\mathbb{C}P^{N-1}$ are given by
\begin{equation}
   {\mit\Gamma}^a{}_{bc}\equiv G^{\Bar{a}a}\partial_b G_{c\Bar{a}},\qquad
   {\mit\Gamma}^{\Bar{a}}{}_{\Bar{b}\Bar{c}}
   \equiv G^{\Bar{a}a}\partial_{\Bar{b}}G_{a\Bar{c}},\qquad
   R^{\Bar{a}}{}_{\Bar{b}c\Bar{d}}
   \equiv\partial_c\mit{\Gamma}^{\Bar{a}}{}_{\Bar{d}\Bar{b}}.
\label{eq:(2.3)}
\end{equation}
In~Eq.~\eqref{eq:(2.1)}, the spacetime derivatives are denoted as
\begin{equation}
   \partial\equiv\frac{1}{2}(\partial_x-i\partial_y),\qquad
   \Bar{\partial}\equiv\frac{1}{2}(\partial_x+i\partial_y),
\label{eq:(2.4)}
\end{equation}
and the topological term~$S_{\text{top}}$ is defined by
\begin{equation}
   S_{\text{top}}
   \equiv\int d^2x\,\frac{i\theta}{\pi}G_{a\Bar{b}}
   \left(
   \partial\varphi^a\Bar{\partial}\Bar{\varphi}^{\Bar{b}}
   -\Bar{\partial}\varphi^a\partial\Bar{\varphi}^{\Bar{b}}
   \right).
   \label{eq:(2.5)}
\end{equation}

Now, along~$S^1$, we impose the following $\mathbb{Z}_N$ invariant twisted
boundary conditions:
\begin{align}
   &\varphi^a(x,y+2\pi R)=e^{2\pi im_aR}\varphi^a(x,y),
\notag\\
   &\psi_\pm^a(x,y+2\pi R)=e^{2\pi im_aR}\psi_\pm^a(x,y),\qquad
   \Bar{\psi}_\pm^a(x,y+2\pi R)
   =e^{-2\pi im_aR}\Bar{\psi}_\pm^a(x,y),
\label{eq:(2.6)}
\end{align}
where the twist angles are proportional to the index~$a$:
\begin{equation}
   m_a=\frac{a}{N}\frac{1}{R},\qquad a=1,\dotsc,N-1.
\label{eq:(2.7)}
\end{equation}
These boundary conditions allow the so-called fractional instanton with a
particular index~$b$ ($b=1$, $2$, \dots, $N-1$):
\begin{equation}
   \varphi^{a\neq b}=0,\qquad
   \varphi^b=\mathcal{C}e^{m_b(x+iy)},\qquad
   \mathcal{C}\in\mathbb{C},
\label{eq:(2.8)}
\end{equation}
which has the classical action
\begin{equation}
   S=2\pi i\left(\frac{\theta}{2\pi}-i\frac{N}{\lambda}\right)m_bR
   =\left(\frac{2\pi}{\lambda}+i\frac{\theta}{N}\right)b.
\label{eq:(2.9)}
\end{equation}
Thus the action of a pair of the $b$th fractional instanton and the $b$th
fractional anti-instanton---called the $b$th bion---approaches
\begin{equation}
   S\sim4\pi\frac{N}{\lambda}m_bR=\frac{4\pi}{\lambda}b,
\label{eq:(2.10)}
\end{equation}
as the separation between the fractional instanton and the fractional
anti-instanton goes to infinity. Since the $b$th bion possesses the
action~$S\sim4\pi b/\lambda$, this would produce a singularity for the Borel
transform $B(u)$ in~Eq.~\eqref{eq:(1.5)} (of a quantity in the topologically
trivial sector) at the value of the classical
action~\cite{Brezin:1976wa,Lipatov:1976ny,LeGuillou:1990nq}, i.e., $u=b$,
where $b=1$, $2$, \dots, $N-1$.

\subsection{Homogeneous coordinates}
\label{sec:2.2}
For the large-$N$ approximation, it is highly convenient to express the above
system in terms of the homogeneous coordinate of~$\mathbb{C}P^{N-1}$. That is,
introducing a new variable~$z^N\in\mathbb{C}$, we set
\begin{equation}
   \varphi^a\equiv\frac{z^a}{z^N},\qquad
   \Bar{z}^Az^A=1.
\label{eq:(2.11)}
\end{equation}
Here and in what follows, uppercase indices $A$, $B$, etc.\ run over $1$,
\dots, $N$. We denote fields with the indices $A$, $B$, \dots\ as $N$~fields.
Note that this description of the system in terms of~$z^A$ is redundant; i.e.,
the original variables $\varphi^a$ are invariant under the $U(1)$ gauge
transformation,
\begin{equation}
   z^A\to gz^A,\qquad g\in U(1).
\label{eq:(2.12)}
\end{equation}

For the fermionic fields, we find that the following variables work quite well
(see also~Ref.~\cite{Cremmer:1978bh}). We introduce new variables~$\chi_\pm^A$
and set
\begin{equation}
   \psi_\pm^a\equiv\frac{1}{z^N}\chi_\pm^a-\frac{z^a}{(z^N)^2}\chi_\pm^N.
\label{eq:(2.13)}
\end{equation}
The original variables $\psi_\pm^a$ are invariant under the $U(1)$ gauge
transformation, defined by the combination of~Eq.~\eqref{eq:(2.12)} and
\begin{equation}
   \chi_\pm^A\to g\chi_\pm^A,\qquad g\in U(1).
\label{eq:(2.14)}
\end{equation}
We then impose the constraint
\begin{equation}
   \Bar{z}^A\chi_\pm^A=0.
\label{eq:(2.15)}
\end{equation}
This constraint has the solution with respect to~$\chi_\pm^N$:
\begin{equation}
   \chi_\pm^N=-(z^N)^2\Bar{z}^a\psi_\pm^a,
\label{eq:(2.16)}
\end{equation}
under the condition~$\Bar{z}^Az^A=1$.

We may assume the following boundary conditions for the homogeneous coordinate
variables:
\begin{align}
   &z^A(x,y+2\pi R)=e^{2\pi im_AR}z^A(x,y),
\notag\\
   &\chi_\pm^A(x,y+2\pi R)=e^{2\pi im_AR}\chi_\pm^A(x,y),\qquad
   \Bar{\chi}_\pm^A(x,y+2\pi R)
   =e^{-2\pi im_AR}\Bar{\chi}_\pm^A(x,y),
\label{eq:(2.17)}
\end{align}
where we have defined
\begin{equation}
   m_N\equiv0,
\label{eq:(2.18)}
\end{equation}
such that these conditions are consistent with~Eq.~\eqref{eq:(2.6)}.

Then, using the basic formulas given in~Appendix~\ref{sec:A}, we find a rather
simple expression for the action:
\begin{align}
   S&=\int d^2x\,\frac{2N}{\lambda}
   \biggl[
   \partial\Bar{z}^A\Bar{\partial}z^A
   +\Bar{\partial}\Bar{z}^A\partial z^A
   -2j_zj_{\Bar{z}}
\notag\\
   &\qquad\qquad\qquad\qquad{}
   +\Bar{\chi}_-^A(\partial-ij_z)\chi_-^A
   +\Bar{\chi}_+^A(\Bar{\partial}-ij_{\Bar{z}})\chi_+^A
\notag\\
   &\qquad\qquad\qquad\qquad{}
   -\frac{1}{2}\chi_+^A\Bar{\chi}_-^A\chi_-^B\Bar{\chi}_+^B
   +\frac{1}{2}\chi_+^A\Bar{\chi}_+^A\chi_-^B\Bar{\chi}_-^B
   \biggr]
\notag\\
   &\qquad{}
   +S_{\text{top}},
\label{eq:(2.19)}
\end{align}
where
\begin{equation}
   j_z=\frac{1}{2i}(\Bar{z}^A\partial z^A-z^A\partial\Bar{z}^A),\qquad
   j_{\Bar{z}}=\frac{1}{2i}(\Bar{z}^A\Bar{\partial}z^A
   -z^A\Bar{\partial}\Bar{z}^A),
\label{eq:(2.20)}
\end{equation}
and
\begin{equation}
   S_{\text{top}}
   =\int d^2x\,
   \frac{\theta}{\pi}\left(\partial j_{\Bar{z}}-\Bar{\partial}j_z\right).
\label{eq:(2.21)}
\end{equation}
This is basically the action given in~Eq.~(15) of~Ref.~\cite{DAdda:1978dle}.
Note that under the $U(1)$ gauge transformation~\eqref{eq:(2.12)}, the
current~\eqref{eq:(2.20)} transforms inhomogeneously:
\begin{equation}
   j_z\to j_z+\frac{1}{i}g^{-1}\partial g,\qquad
   j_{\Bar{z}}\to j_{\Bar{z}}+\frac{1}{i}g^{-1}\Bar{\partial}g.
\label{eq:(2.22)}
\end{equation}

\subsection{Auxiliary fields}
We now introduce various auxiliary fields. One of their roles is to impose the
constraint in~Eq.~\eqref{eq:(2.11)} and the fermionic
constraint~\eqref{eq:(2.15)}; the corresponding Lagrange multiplier fields are
$f$ and $(\eta_\pm,\Bar{\eta}_\pm)$, respectively. We also introduce auxiliary
fields $A_{z,\Bar{z}}$ and~$(\sigma,\Bar{\sigma})$ to make the action quadratic
in the homogeneous coordinate variables. We thus set
\begin{align}
   S'&\equiv
   S+\int d^2x\,\frac{2N}{\lambda}
   \biggl[
   \frac{1}{2}f(\Bar{z}^Az^A-1)
   +\Bar{\eta}_-\Bar{z}^A\chi_+^A
   +\Bar{\eta}_+\Bar{z}^A\chi_-^A
   +\Bar{\chi}_+^Az^A\eta_-
   +\Bar{\chi}_-^Az^A\eta_+
\notag\\
   &\qquad\qquad\qquad\qquad{}
   +2\left(A_{\Bar{z}}+j_{\Bar{z}}+\frac{1}{2}i\Bar{\chi}_-^A\chi_-^A\right)
   \left(A_z+j_z+\frac{1}{2}i\Bar{\chi}_+^B\chi_+^B\right)
\notag\\
   &\qquad\qquad\qquad\qquad{}
   +\frac{1}{2}
   (\Bar{\sigma}+\Bar{\chi}_+^A\chi_-^A)
   (\sigma+\Bar{\chi}_-^B\chi_+^B)
   \biggr]
\notag\\
   &\qquad{}
   -\int d^2x\,\frac{\theta}{\pi}
   \left[
   \partial
   \left(A_{\Bar{z}}+j_{\Bar{z}}+\frac{1}{2}i\Bar{\chi}_-^A\chi_-^A\right)
   -\Bar{\partial}
   \left(A_z+j_z+\frac{1}{2}i\Bar{\chi}_+^A\chi_+^A\right)
   \right].
\label{eq:(2.23)}
\end{align}
We impose the periodic boundary conditions for all the auxiliary fields. The
action~$S'$ can be cast into the form
\begin{align}
   S'
   &=\int d^2x\,\frac{N}{\lambda}\Biggl\{-f+\Bar{\sigma}\sigma
   +\Bar{z}^A
   \left[
   -D_\mu D_\mu+f
   -4\Bar{\eta}
   (\Slash{D}+\Bar{\sigma}P_++\sigma P_-)^{-1}
   \eta\right]
   z^A
\notag\\
   &\qquad\qquad\qquad{}
   +(\widetilde{\Bar{\chi}}_-^A\,\widetilde{\Bar{\chi}}_+^A)
   (\Slash{D}+\Bar{\sigma}P_++\sigma P_-)
   \begin{pmatrix}\widetilde{\chi}_+^A\\\widetilde{\chi}_-^A\end{pmatrix}
   \Biggr\}
\notag\\
   &\qquad{}
   -\int d^2x\,\frac{i\theta}{2\pi}\epsilon_{\mu\nu}\partial_\mu A_\nu,
\label{eq:(2.24)}
\end{align}
where, setting $A_x=A_z+A_{\Bar{z}}$ and~$A_y=i(A_z-A_{\Bar{z}})$,
\begin{equation}
   D_\mu z^A\equiv(\partial_\mu+iA_\mu)z^A,\qquad
   \Slash{D}\begin{pmatrix}\chi_+^A\\\chi_-^A\end{pmatrix}
   \equiv\gamma_\mu(\partial_\mu+iA_\mu)
   \begin{pmatrix}\chi_+^A\\\chi_-^A\end{pmatrix},
\label{eq:(2.25)}
\end{equation}
\begin{equation}
   \gamma_x\equiv\begin{pmatrix}0&1\\1&0\\\end{pmatrix},\qquad
   \gamma_y\equiv\begin{pmatrix}0&-i\\i&0\\\end{pmatrix},\qquad
   P_\pm\equiv\frac{1\pm\gamma_5}{2},\qquad
   \gamma_5\equiv-i\gamma_x\gamma_y,
\label{eq:(2.26)}
\end{equation}
and
\begin{equation}
   \Bar{\eta}\equiv(\Bar{\eta}_-\,\Bar{\eta}_+),\qquad
   \eta\equiv\begin{pmatrix}\eta_+\\\eta_-\end{pmatrix},
\label{eq:(2.27)}
\end{equation}
and $\epsilon_{xy}=-\epsilon_{yx}=+1$. Also, we have defined
\begin{align}
   (\widetilde{\Bar{\chi}}_-^A\,\widetilde{\Bar{\chi}}_+^A)
   &\equiv
   (\Bar{\chi}_-^A\,\Bar{\chi}_+^A)
   +2\Bar{\eta}\Bar{z}^A
   (\Slash{D}+\Bar{\sigma}P_++\sigma P_-)^{-1},
\notag\\
   \begin{pmatrix}\widetilde{\chi}_+^A\\\widetilde{\chi}_-^A\end{pmatrix}
   &\equiv
   \begin{pmatrix}\chi_+^A\\\chi_-^A\end{pmatrix}
   +2(\Slash{D}+\Bar{\sigma}P_++\sigma P_-)^{-1}
   z^A\eta.
\label{eq:(2.28)}
\end{align}

Since the current~\eqref{eq:(2.20)} transforms inhomogeneously under the $U(1)$
gauge transformation as~Eq.~\eqref{eq:(2.22)}, the auxiliary field~$A_\mu$
receives a $U(1)$ gauge transformation of the form
\begin{equation}
   A_\mu\to A_\mu-\frac{1}{i}g^{-1}\partial_\mu g,
\label{eq:(2.29)}
\end{equation}
in order for the term added in~Eq.~\eqref{eq:(2.23)} to be gauge invariant.
$A_\mu$ is therefore regarded as a $U(1)$ gauge potential and the last term
of~Eq.~\eqref{eq:(2.24)} can give rise to a non-trivial topological charge.

\section{Leading-order large-$N$ approximation}
\label{sec:3}
\subsection{The saddle point}
The large-$N$ approximation consists of the saddle point approximation of the
functional integral of auxiliary fields, after the Gaussian integration over
the original $N$~fields~\cite{Coleman:1985rnk}. From~Eq.~\eqref{eq:(2.24)}, the
integration over the $N$~fields yields the effective action of the auxiliary
fields:
\begin{align}
   S_{\text{eff}}&=\int d^2x\,\frac{N}{\lambda}(-f+\Bar{\sigma}\sigma)
\notag\\
   &\qquad{}
   +\sum_A\Tr\Ln\left[
   -D_\mu D_\mu+f
   -4\Bar{\eta}(\Slash{D}+\Bar{\sigma}P_++\sigma P_-)^{-1}\eta
   \right]
\notag\\
   &\qquad{}
   -\sum_A\Tr\Ln(\Slash{D}+\Bar{\sigma}P_++\sigma P_-).
\label{eq:(3.1)}
\end{align}
In this expression, the twisted boundary conditions in~Eq.~\eqref{eq:(2.17)}
that depend on the index~$A$ have to be taken into account.

First, we look for the saddle point of the effective action, by assuming that
it is given by\footnote{For the uncompactified space~$\mathbb{R}^2$, we should
set~$A_{\mu0}=0$ for the Lorentz invariance and we do not have the integration
over~$A_{y0}$ in~Eq.~\eqref{eq:(3.24)}.}
\begin{equation}
   A_{\mu0}=\text{const.},\qquad
   f_0=\text{const.},\qquad
   \sigma_0=\text{const.},\qquad
   \eta_0=\Bar{\eta}_0=0.
\label{eq:(3.2)}
\end{equation}
For such a constant configuration, one can see that
\begin{align}
   \Tr\Ln(\Slash{D}+\Bar{\sigma}P_++\sigma P_-)
   =\Tr\Ln(-D_\mu D_\mu+\Bar{\sigma}\sigma),
\label{eq:(3.3)}
\end{align}
by using the charge conjugation invariance. Then, for
the configuration~\eqref{eq:(3.2)}, going to the momentum space by taking the
twisted boundary conditions~\eqref{eq:(2.17)} into account, we have
\begin{align}
   S_{\text{eff}}
   &=\int d^2x\,\frac{N}{\lambda}(-f_0+\Bar{\sigma}_0\sigma_0)
\notag\\
   &\qquad{}
   +\sum_A\int d^2x\,\int\frac{dp_x}{2\pi}\frac{1}{2\pi R}\sum_{p_y}
   \ln\left[(p_x+A_{x0})^2+(p_y+m_A+A_{y0})^2+f_0\right]
\notag\\
   &\qquad{}
   -\sum_A\int d^2x\,\int\frac{dp_x}{2\pi}\frac{1}{2\pi R}\sum_{p_y}
   \ln\left[
   (p_x+A_{x0})^2+(p_y+m_A+A_{y0})^2+\Bar{\sigma}_0\sigma_0
   \right],
\label{eq:(3.4)}
\end{align}
where the KK momentum $p_y$ is discrete:
\begin{equation}
   p_y=\frac{n}{R},\qquad n\in\mathbb{Z}.
\label{eq:(3.5)}
\end{equation}
Then, we use the identity
\begin{equation}
   \sum_{n=-\infty}^\infty e^{ip_y2\pi Rn}
   =\frac{1}{R}\sum_{n=-\infty}^\infty\delta(p_y-n/R)
\label{eq:(3.6)}
\end{equation}
or
\begin{equation}
   \frac{1}{2\pi R}\sum_{n=-\infty}^\infty F(n/R)
   =\sum_{n=-\infty}^\infty\int\frac{dp_y}{2\pi}\,e^{ip_y2\pi Rn}F(p_y)
\label{eq:(3.7)}
\end{equation}
in~Eq.~\eqref{eq:(3.4)} to make the sum~$\sum_{p_y}$ into integrals~$\int dp_y$.
This enables us to shift the integration variables as $p_x\to p_x-A_{x0}$
and~$p_x\to p_x-m_A-A_{x0}$ to yield
\begin{align}
   S_{\text{eff}}
   &=\int d^2x\,\frac{N}{\lambda}(-f_0+\Bar{\sigma}_0\sigma_0)
\notag\\
   &\qquad{}
   +\sum_A\int d^2x\,\sum_{n=-\infty}^\infty
   \int\frac{d^2p}{(2\pi)^2}\,e^{i(p_y-A_{y0}-m_A)2\pi Rn}
   \left[\ln(p^2+f_0)-\ln(p^2+\Bar{\sigma}_0\sigma_0)\right].
\label{eq:(3.8)}
\end{align}
We then carry out the sum over~$A$ by noting, from~Eqs.~\eqref{eq:(2.7)}
and~\eqref{eq:(2.18)},
\begin{equation}
   \sum_Ae^{-im_A2\pi Rn}=\sum_{j=0}^{N-1}\left(e^{-2\pi ni/N}\right)^j
   =\begin{cases}
   N&\text{for $n=0\bmod N$},\\
   0&\text{for $n\neq0\bmod N$}.\\
   \end{cases}
\label{eq:(3.9)}
\end{equation}
We thus obtain
\begin{align}
   S_{\text{eff}}
   &=\int d^2x\,\frac{N}{\lambda}(-f_0+\Bar{\sigma}_0\sigma_0)
\notag\\
   &\qquad{}
   +\int d^2x\,N\sum_{m=-\infty}^\infty
   \int\frac{d^2p}{(2\pi)^2}\,e^{i(p_y-A_{y0})2\pi RNm}
   \left[\ln(p^2+f_0)-\ln(p^2+\Bar{\sigma}_0\sigma_0)\right].
\label{eq:(3.10)}
\end{align}
In this form, the $m=0$ term is ultraviolet (UV) divergent whereas $m\neq0$
terms are the Fourier transforms and UV finite. To the $m=0$ term, we apply
dimensional regularization where the dimension of spacetime is set to
be~$2\to D\equiv2-2\varepsilon$. The result of the momentum integrations is
then
\begin{align}
   S_{\text{eff}}
   &=\int d^2x\,
   \frac{N}{4\pi}\left[
   \frac{4\pi}{\lambda}
   -\frac{1}{\varepsilon}+\ln\left(\frac{e^{\gamma_E}}{4\pi}\right)\right]
   (-f_0+\Bar{\sigma}_0\sigma_0)
\notag\\
   &\qquad{}
   +\int d^2x\,\frac{N}{4\pi}
   \left\{
   -f_0(\ln f_0-1)+\Bar{\sigma}_0\sigma_0
   \left[\ln(\Bar{\sigma}_0\sigma_0)-1\right]
   \right\}
\notag\\
   &\qquad{}
   +\int d^2x\,\frac{N}{4\pi}
   (-4)\sum_{m\neq0}e^{-iA_{y0}2\pi RNm}
\notag\\
   &\qquad\qquad{}
   \times
   \left[
   \frac{\sqrt{f_0}}{2\pi RN|m|}K_1(\sqrt{f_0}2\pi RN|m|)
   -\frac{\sqrt{\Bar{\sigma}_0\sigma_0}}{2\pi RN|m|}
   K_1(\sqrt{\Bar{\sigma}_0\sigma_0}2\pi RN|m|)
   \right].
\label{eq:(3.11)}
\end{align}
Here and in what follows, $K_\nu(z)$ denotes the modified Bessel function of
the second kind. To remove the UV divergence in~Eq.~\eqref{eq:(3.11)}, we
introduce the renormalized 't~Hooft coupling~$\lambda_R(\mu)$ in the
``$\overline{\text{MS}}$ scheme'' by
\begin{equation}
   \lambda=\left(\frac{e^{\gamma_E}\mu^2}{4\pi}\right)^\varepsilon\lambda_R(\mu)
   \left[1+\frac{\lambda_R(\mu)}{4\pi}\frac{1}{\varepsilon}\right]^{-1},
\label{eq:(3.12)}
\end{equation}
where $\mu$ is a renormalization scale. From this result, we obtain the beta
function
\begin{equation}
   \left.\mu\frac{\partial}{\partial\mu}\lambda_R(\mu)\right|_\lambda
   =-2\varepsilon\lambda_R(\mu)-\frac{1}{2\pi}\lambda_R(\mu)^2,
\label{eq:(3.13)}
\end{equation}
and the renormalization-group invariant dynamical scale (the
$\Lambda$~parameter)
\begin{equation}
   \Lambda\equiv\mu e^{-2\pi/\lambda_R(\mu)}.
\label{eq:(3.14)}
\end{equation}

In terms of~$\Lambda$, the effective action~\eqref{eq:(3.11)} reads
\begin{equation}
   S_{\text{eff}}
   =\int d^2x\,\frac{N}{4\pi}\left[V(f_0)-V(\Bar{\sigma}_0\sigma_0)\right],
\label{eq:(3.15)}
\end{equation}
where the function $V(z)$ is defined by
\begin{equation}
   V(z)\equiv V_\infty(z)+\Hat{V}(z),
\label{eq:(3.16)}
\end{equation}
with
\begin{equation}
   V_\infty(z)\equiv-z\left[\ln(z/\Lambda^2)-1\right],
\label{eq:(3.17)}
\end{equation}
and
\begin{equation}
   \Hat{V}(z)\equiv
   -4\sum_{m\neq0}e^{-iA_{y0}2\pi RNm}
   \frac{\sqrt{z}}{2\pi RN|m|}K_1(\sqrt{z}2\pi RN|m|).
\label{eq:(3.18)}
\end{equation}
The infinite sum in~$\Hat{V}(z)$ is convergent because
$K_\nu(z)\stackrel{z\to\infty}{\sim}\sqrt{\pi/(2z)}e^{-z}$. Moreover, as shown
in~Appendix~\ref{sec:B}, $\Hat{V}(z)\to0$ in the large-$N$
limit~\eqref{eq:(1.1)}. Therefore, as~$N\to\infty$, $f_0$
and~$\Bar{\sigma}_0\sigma_0$ are given by the solution of
\begin{equation}
   V_\infty'(z)=-\ln(z/\Lambda^2)=0,
\label{eq:(3.19)}
\end{equation}
i.e.,
\begin{equation}
   f_0=\Bar{\sigma}_0\sigma_0=\Lambda^2.
\label{eq:(3.20)}
\end{equation}
These are identical to the values in the system
in~$\mathbb{R}^2$~\cite{DAdda:1978dle}.

At the above saddle point, from~Eq.~\eqref{eq:(3.15)}, $S_{\text{eff}}\equiv0$
and $S_{\text{eff}}$ becomes independent of~$A_{y0}$. Therefore, $A_{y0}$ is not
determined from the saddle point condition in the present supersymmetric system
on~$\mathbb{R}\times S^1$. We should perform the integration over this
``vacuum moduli'' $A_{y0}$ in the functional integral. We note that the
action~\eqref{eq:(2.24)} and the boundary conditions~\eqref{eq:(2.17)} are
invariant under the ``center transformation'' defined by
\begin{equation}
   z^A\to gz^A,\qquad
   \chi_\pm^A\to g\chi_\pm^A,
\label{eq:(3.21)}
\end{equation}
with~Eq.~\eqref{eq:(2.29)}, where $g\in U(1)$ obeys the non-trivial boundary
condition,
\begin{equation}
   g(x,y+2\pi R)=e^{2\pi i/N}g(x,y),
\label{eq:(3.22)}
\end{equation}
if we relabel $N$~fields as $z^A\to z^{\Tilde{A}}$
and~$\chi_\pm^A\to\chi_\pm^{\Tilde{A}}$, where $\Tilde{A}=A+1$
for~$1\leq A\leq N-1$ and $\Tilde{A}=1$ for~$A=N$. An element $g=e^{iy/(RN)}$ of
this center transformation induces a constant shift on~$A_{y0}$
through~Eq.~\eqref{eq:(2.29)}:
\begin{equation}
   A_{y0}\to A_{y0}-\frac{1}{RN}.
\label{eq:(3.23)}
\end{equation}
Hence, for gauge invariant quantities that are invariant under the relabeling,
$z^A\to z^{\Tilde{A}}$ and~$\chi_\pm^A\to\chi_\pm^{\Tilde{A}}$, such as the
partition function and the gluon condensate considered below, the integration
over~$A_{y0}$ should be restricted in the ``fundamental domain'' as
\begin{equation}
   \int_0^1d(A_{y0}RN).
\label{eq:(3.24)}
\end{equation}

\subsection{Effective action for fluctuations}
We next compute the effective action for fluctuations of the auxiliary fields
around the above large-$N$ saddle point. That is, setting
\begin{equation}
   A_\mu\equiv A_{\mu0}+\delta A_\mu,\qquad
   f\equiv f_0+\delta f,\qquad
   \sigma\equiv\sigma_0+\delta\sigma,
\label{eq:(3.25)}
\end{equation}
we compute $S_{\text{eff}}$ to the quadratic order in the fluctuations.

For illustration, let us consider
\begin{equation}
   \sum_A\left.\Tr\Ln\left[
   -D_\mu D_\mu+f
   -4\Bar{\eta}(\Slash{D}+\Bar{\sigma}P_++\sigma P_-)^{-1}\eta
   \right]\right|_{O(\delta f^2)}.
\label{eq:(3.26)}
\end{equation}
Going to the momentum space and using the relations~\eqref{eq:(3.7)}
and~\eqref{eq:(3.9)}, we have
\begin{align}
   &\sum_A\left.\Tr\Ln\left[
   -D_\mu D_\mu+f
   -4\Bar{\eta}(\Slash{D}+\Bar{\sigma}P_++\sigma P_-)^{-1}\eta
   \right]\right|_{O(\delta f^2)}
\notag\\
   &=N\int\frac{dp_x}{2\pi}\,\frac{1}{2\pi R}\sum_{p_y}
   \left(-\frac{1}{2}\right)\widetilde{\delta f}(p)\widetilde{\delta f}(-p)
   \sum_{m=-\infty}^\infty e^{-iA_{y0}2\pi RNm}
\notag\\
   &\qquad{}
   \times\int_0^1 dx\,\int\frac{d^2k}{(2\pi)^2}\,
   e^{ik_y2\pi RNm}
   \frac{1}{(k^2-2xkp+f_0+xp^2)^2},
\label{eq:(3.27)}
\end{align}
where we have set
\begin{equation}
   \delta f(x)\equiv
   \int\frac{dp_x}{2\pi}\,\frac{1}{2\pi R}\sum_{p_y}
   e^{ipx}\widetilde{\delta f}(p).
\label{eq:(3.28)}
\end{equation}
The momentum integration then yields
\begin{align}
   &\sum_A\left.\Tr\Ln\left[
   -D_\mu D_\mu+f
   -4\Bar{\eta}(\Slash{D}+\Bar{\sigma}P_++\sigma P_-)^{-1}\eta
   \right]\right|_{O(\delta f^2)}
\notag\\
   &=\frac{N}{4\pi}\int\frac{dp_x}{2\pi}\,\frac{1}{2\pi R}\sum_{p_y}
   \left(-\frac{1}{2}\right)\widetilde{\delta f}(p)\widetilde{\delta f}(-p)
   \mathcal{L}(p).
\label{eq:(3.29)}
\end{align}
Here, we have introduced the function
\begin{equation}
   \mathcal{L}(p)\equiv\mathcal{L}_\infty(p)+\Hat{\mathcal{L}}(p),
\label{eq:(3.30)}
\end{equation}
where (using $f_0=\Lambda^2$~\eqref{eq:(3.20)})
\begin{equation}
   \mathcal{L}_\infty(p)
   \equiv\frac{2}{\sqrt{p^2(p^2+4\Lambda^2)}}
   \ln\left(\frac{\sqrt{p^2+4\Lambda^2}+\sqrt{p^2}}
   {\sqrt{p^2+4\Lambda^2}-\sqrt{p^2}}\right)
\label{eq:(3.31)}
\end{equation}
is the expression common to the uncompactifed space~$\mathbb{R}^2$, and
\begin{align}
   \Hat{\mathcal{L}}(p)&\equiv
   \int_0^1dx\,
   \sum_{m\neq0}e^{-iA_{y0}2\pi RNm}e^{ixp_y2\pi RNm}
\notag\\
   &\qquad\qquad\qquad{}
   \times\frac{2\pi RN|m|}{\sqrt{\Lambda^2+x(1-x)p^2}}
   K_1(\sqrt{\Lambda^2+x(1-x)p^2}2\pi RN|m|)
\label{eq:(3.32)}
\end{align}
is a part peculiar to the compactified space~$\mathbb{R}\times S^1$. Note that
$\mathcal{L}_\infty(p)$ and~$\Hat{\mathcal{L}}(p)$ are real and
\begin{equation}
   \mathcal{L}_\infty(-p)=\mathcal{L}_\infty(p),\qquad
   \Hat{\mathcal{L}}(-p)=\Hat{\mathcal{L}}(p).
\label{eq:(3.33)}
\end{equation}
To show the latter property, we note that the change
$e^{ixp_y2\pi RNm}\to e^{-ixp_y2\pi RNm}=e^{i(1-x)p_y2\pi RNm}$ caused by~$p\to-p$ can
be absorbed by the change of the integration variable $x\to1-x$ (recall that
$p_y=n/R$ with~$n\in\mathbb{Z}$).

Repeating similar calculations by setting
\begin{align}
   \delta A_\mu(x)
   &\equiv\int\frac{dp_x}{2\pi}\,\frac{1}{2\pi R}\sum_{p_y}
   e^{ipx}\widetilde{\delta A}_\mu(p),&
   \delta\sigma(x)
   &\equiv\int\frac{dp_x}{2\pi}\,\frac{1}{2\pi R}\sum_{p_y}
   e^{ipx}\widetilde{\delta\sigma}(p),
\notag\\
   \eta(x)
   &\equiv\int\frac{dp_x}{2\pi}\,\frac{1}{2\pi R}\sum_{p_y}
   e^{ipx}\widetilde{\eta}(p),&
   \Bar{\eta}(x)
   &\equiv\int\frac{dp_x}{2\pi}\,\frac{1}{2\pi R}\sum_{p_y}
   e^{ipx}\widetilde{\Bar{\eta}}(p),
\label{eq:(3.34)}
\end{align}
and
\begin{equation}
   \widetilde{\delta R}(p)
   \equiv\frac{1}{2}\left[
   \Bar{\sigma}_0\widetilde{\delta\sigma}(p)
   +\sigma_0\widetilde{\delta\Bar{\sigma}}(p)
   \right],\qquad
   \widetilde{\delta I}(p)
   \equiv\frac{1}{2i}\left[
   \Bar{\sigma}_0\widetilde{\delta\sigma}(p)
   -\sigma_0\widetilde{\delta\Bar{\sigma}}(p)
   \right],
\label{eq:(3.35)}
\end{equation}
we obtain\footnote{To obtain this simplified form, we have to do integration by
parts with respect to the Feynman parameter~$x$ by using relations such as
$K_0'(z)=-K_1(z)$ and~$zK_1'(z)+K_1(z)=-zK_0(z)$. Also, we have defined the
$\gamma_5$ in dimensional regularization such that
$\gamma_5\equiv-i\gamma_x\gamma_y$ for any~$D$~\cite{tHooft:1972tcz}; thus it
commutes with~$\gamma_\mu$ when $\mu\neq x$ or~$\mu\neq y$.
\label{footnote:4}}
\begin{align}
   &S_{\text{eff}}|_{\text{quadratic}}
\notag\\
   &=\frac{N}{4\pi}\int\frac{dp_x}{2\pi}\,\frac{1}{2\pi R}\sum_{p_y}
\notag\\
   &\qquad\qquad{}
   \times\biggl(
   \frac{1}{2}(p^2\delta_{\mu\nu}-p_\mu p_\nu)\mathcal{L}(p)
   \widetilde{\delta A}_\mu(p)\widetilde{\delta A}_\nu(-p)
\notag\\
   &\qquad\qquad\qquad{}
   +\frac{1}{2\Lambda^2}(p^2+4\Lambda^2)\mathcal{L}(p)
   \widetilde{\delta R}(p)\widetilde{\delta R}(-p)
   +\frac{1}{2\Lambda^2}p^2\mathcal{L}(p)
   \widetilde{\delta I}(p)\widetilde{\delta I}(-p)
\notag\\
   &\qquad\qquad\qquad{}
   -\frac{1}{2}\mathcal{L}(p)\widetilde{\delta f}(p)\widetilde{\delta f}(-p)
\notag\\
   &\qquad\qquad\qquad{}
   -2\Tilde{\Bar{\eta}}(p)
   (i\Slash{p}+2\sigma_0P_++2\Bar{\sigma}_0P_-)\mathcal{L}(p)
   \Tilde{\eta}(-p)
\notag\\
   &\qquad\qquad\qquad{}
   +\epsilon_{\mu\nu}p_\mu\mathcal{L}(p)\widetilde{\delta A}_\nu(p)
   \widetilde{\delta I}(-p)
   -\epsilon_{\mu\nu}p_\mu\mathcal{L}(p)
   \widetilde{\delta I}(p)\widetilde{\delta A}_\nu(-p)
\notag\\
   &\qquad\qquad\qquad{}
   +\left(\delta_{\mu y}-\frac{p_\mu p_y}{p^2}\right)\mathcal{K}(p)
\notag\\
   &\qquad\qquad\qquad\qquad{}
   \times\left\{
   \widetilde{\delta A_\mu}(p)
   \left[2\widetilde{\delta R}(-p)-\widetilde{\delta f}(-p)\right]
   +\left[2\widetilde{\delta R}(p)-\widetilde{\delta f}(p)\right]
   \widetilde{\delta A_\mu}(-p)
   \right\}
\notag\\
   &\qquad\qquad\qquad{}
   -\frac{1}{\Lambda^2}\epsilon_{\mu y}p_\mu\mathcal{K}(p)
   \left[\widetilde{\delta R}(p)\widetilde{\delta I}(-p)
   -\widetilde{\delta I}(p)\widetilde{\delta R}(-p)\right]
\notag\\
   &\qquad\qquad\qquad{}
   +4i\left(\delta_{\mu y}-\frac{p_\mu p_y}{p^2}\right)\mathcal{K}(p)
   \Tilde{\Bar{\eta}}(p)\gamma_\mu
   \Tilde{\eta}(-p)
   \biggr)
\notag\\
   &\qquad+S_{\text{local}},
\label{eq:(3.36)}
\end{align}
where we have used the fact that $\Bar{\sigma}_0\sigma_0=\Lambda^2$
(recall~Eq.~\eqref{eq:(3.20)}) and introduced another function:
\begin{equation}
   \mathcal{K}(p)\equiv
   i\int_0^1dx\,\sum_{m\neq0}e^{-iA_{y0}2\pi RNm}e^{ixp_y2\pi RNm}2\pi RNm
   K_0(\sqrt{\Lambda^2+x(1-x)p^2}2\pi RN|m|).
\label{eq:(3.37)}
\end{equation}
Note that $\mathcal{K}(p)$ is real and
\begin{equation}
   \mathcal{K}(-p)=\mathcal{K}(p).
\label{eq:(3.38)}
\end{equation}
The last term of~Eq.~\eqref{eq:(3.36)} is given by
\begin{align}
   S_{\text{local}}
   &\equiv
   \frac{N}{4\pi}\int\frac{dp_x}{2\pi}\,\frac{1}{2\pi R}\sum_{p_y}\Biggl\{
   \frac{1}{2}
   \left[
   \widetilde{\delta\sigma}(p)\widetilde{\delta\sigma}(-p)
   +\widetilde{\delta\Bar{\sigma}}(p)
   \widetilde{\delta\Bar{\sigma}}(-p)
   \right]
\notag\\
   &\qquad\qquad\qquad\qquad{}
   -\widetilde{\delta\Bar{\sigma}}(p)\widetilde{\delta\sigma}(-p)
   \Biggl[1+2\sum_{m\neq0}
   e^{-iA_{y0}2\pi RNm}K_0(\Lambda2\pi RN|m|)\Biggr]
   \Biggr\}.
\label{eq:(3.39)}
\end{align}
This term breaks the $U(1)$ chiral symmetry, the invariance of the classical
action under $\sigma\to e^{2i\alpha}\sigma$,
$\Bar{\sigma}\to e^{-2i\alpha}\Bar{\sigma}$, $\eta\to e^{-i\alpha\gamma_5}\eta$,
and~$\Bar{\eta}\to\Bar{\eta}e^{-i\alpha\gamma_5}$, and may be regarded as a
``quantum anomaly''. However, since this term is local in position space, we
may simply remove this by a local counterterm as an artifact arising from our
particular definition of the $\gamma_5$ matrix in dimensional regularization.
In what follows, we assume this and neglect~$S_{\text{local}}$.

Equation~\eqref{eq:(3.36)} provides the effective action for the fluctuations
around the large-$N$ saddle point; this generalizes Eq.~(58)
of~Ref.~\cite{DAdda:1978dle} to the case of the twisted boundary conditions on
the compactified space~$\mathbbm{R}\times S^1$.

\subsection{Propagators}
To obtain the propagators of the auxiliary fields from~Eq.~\eqref{eq:(3.36)},
we add a gauge-fixing term
\begin{equation}
   S_{\text{gf}}=\frac{N}{4\pi}\int\frac{dp_x}{2\pi}\,\frac{1}{2\pi R}\sum_{p_y}
   \frac{1}{2}p_\mu p_\nu
   \mathcal{L}(p)\widetilde{\delta A_\mu}(p)\widetilde{\delta A_\nu}(-p)
\label{eq:(3.40)}
\end{equation}
to Eq.~\eqref{eq:(3.36)}. After some calculation, we obtain the
$A_\mu$~propagator as
\begin{align}
   &\left\langle\widetilde{\delta A_\mu}(p)\widetilde{\delta A}_\nu(q)
   \right\rangle
\notag\\
   &=\frac{4\pi}{N}\frac{\mathcal{L}(p)}{\mathcal{D}(p)}
   \left\{\delta_{\mu\nu}
   +4\left[\Lambda^2+(1-p_y^2/p^2)\frac{\mathcal{K}(p)^2}{\mathcal{L}(p)^2}
   \right]
   \frac{p_\mu p_\nu}{(p^2)^2}
   \right\}
   2\pi\delta(p_x+q_x)2\pi R\delta_{p_y+q_y,0}.
\label{eq:(3.41)}
\end{align}
For completeness, we list all the propagators:
\begin{align}
   &\left\langle\widetilde{\delta A_\mu}(p)\widetilde{\delta R}(q)
   \right\rangle
   =\left\langle\widetilde{\delta R}(p)\widetilde{\delta A_\mu}(q)
   \right\rangle=0,
\notag\\
   &\left\langle\widetilde{\delta A_\mu}(p)\widetilde{\delta I}(q)
   \right\rangle
   =-\left\langle\widetilde{\delta I}(p)\widetilde{\delta A_\mu}(q)
   \right\rangle
   =\frac{4\pi}{N}\frac{\mathcal{L}(p)}{\mathcal{D}(p)}
   \frac{2\Lambda^2\Bar{p}_\mu}{p^2}\,
   2\pi\delta(p_x+q_x)2\pi R\delta_{p_y+q_y,0},
\notag\\
   &\left\langle\widetilde{\delta A_\mu}(p)\widetilde{\delta f}(q)
   \right\rangle
   =\left\langle\widetilde{\delta f}(p)\widetilde{\delta A_\mu}(q)
   \right\rangle
   =\frac{4\pi}{N}\frac{\mathcal{K}(p)}{\mathcal{D}(p)}
   \frac{-2\Bar{p}_\mu\Bar{p}_y}{p^2}\,
   2\pi\delta(p_x+q_x)2\pi R\delta_{p_y+q_y,0},
\notag\\
   &\left\langle\widetilde{\delta R}(p)\widetilde{\delta R}(q)
   \right\rangle
   =\frac{4\pi}{N}\frac{\mathcal{L}(p)}{\mathcal{D}(p)}\Lambda^2\,
   2\pi\delta(p_x+q_x)2\pi R\delta_{p_y+q_y,0},
\notag\\
   &\left\langle\widetilde{\delta R}(p)\widetilde{\delta I}(q)
   \right\rangle
   =-\left\langle\widetilde{\delta I}(p)\widetilde{\delta R}(q)
   \right\rangle
   =\frac{4\pi}{N}\frac{\mathcal{K}(p)}{\mathcal{D}(p)}
   \frac{-2\Lambda^2\Bar{p}_y}{p^2}\,
   2\pi\delta(p_x+q_x)2\pi R\delta_{p_y+q_y,0},
\notag\\
   &\left\langle\widetilde{\delta R}(p)\widetilde{\delta f}(q)
   \right\rangle
   =\left\langle\widetilde{\delta f}(p)\widetilde{\delta R}(q)
   \right\rangle
   =0,
\notag\\
   &\left\langle\widetilde{\delta I}(p)\widetilde{\delta I}(q)
   \right\rangle
   =\frac{4\pi}{N}\frac{\mathcal{L}(p)}{\mathcal{D}(p)}\Lambda^2\,
   2\pi\delta(p_x+q_x)2\pi R\delta_{p_y+q_y,0},
\notag\\
   &\left\langle\widetilde{\delta I}(p)\widetilde{\delta f}(q)
   \right\rangle
   =-\left\langle\widetilde{\delta f}(p)\widetilde{\delta I}(q)
   \right\rangle
   =\frac{4\pi}{N}\frac{\mathcal{K}(p)}{\mathcal{D}(p)}
   \frac{4\Lambda^2\Bar{p}_y}{p^2}\,
   2\pi\delta(p_x+q_x)2\pi R\delta_{p_y+q_y,0},
\notag\\
   &\left\langle\widetilde{\delta f}(p)\widetilde{\delta f}(q)
   \right\rangle
   =\frac{4\pi}{N}\frac{\mathcal{L}(p)}{\mathcal{D}(p)}(-1)(p^2+4\Lambda^2)\,
   2\pi\delta(p_x+q_x)2\pi R\delta_{p_y+q_y,0},
\label{eq:(3.42)}
\end{align}
and
\begin{align}
   &\left\langle\Tilde{\eta}(p)\Tilde{\Bar{\eta}}(q)\right\rangle
\notag\\
   &=\frac{4\pi}{N}
   \frac{
   (i\Slash{p}+2\Bar{\sigma}_0P_++2\sigma_0P_-)\mathcal{L}(p)
   +2i(\gamma_y-\Slash{p}p_y/p^2)\mathcal{K}(p)}
   {\mathcal{D}(p)}\left(-\frac{1}{2}\right)
   2\pi\delta(p_x+q_x)2\pi R\delta_{p_y+q_y,0}.
\label{eq:(3.43)}
\end{align}
In the above expressions, we have defined
\begin{equation}
   \mathcal{D}(p)
   \equiv(p^2+4\Lambda^2)\mathcal{L}(p)^2+4(1-p_y^2/p^2)\mathcal{K}(p)^2,\qquad
   \Bar{p}_\mu\equiv\epsilon_{\nu\mu}p_\nu,
\label{eq:(3.44)}
\end{equation}
and, in deriving those expressions, we have noted the relation holding in the
two dimensions,
\begin{equation}
   \Bar{p}_\mu\Bar{p}_\nu=p^2\delta_{\mu\nu}-p_\mu p_\nu.
\label{eq:(3.45)}
\end{equation}

\section{IR renormalon in the gluon condensate}
\label{sec:4}
In this section, we compute the gluon condensate in the leading order of the
large-$N$ approximation and extract the perturbative part from it. We then
obtain the corresponding Borel transform~$B(u)$ and study its singularities.

\begin{figure}
\centering
\includegraphics[width=0.2\textwidth,clip]{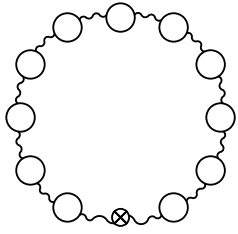}
\caption{The Feynman diagram corresponding to the gluon
condensate~\eqref{eq:(4.2)}. The blob is the combination
in~Eq.~\eqref{eq:(4.1)}. The $A_\mu$ propagator~\eqref{eq:(3.41)} is given by
the chain of the one-loop vacuum polarization diagrams owing to the
$N$~fields.}
\label{fig:1}
\end{figure}

The gluon condensate is given by
\begin{equation}
   \left\langle F_{\mu\nu}(x)F_{\mu\nu}(x)\right\rangle
   =\left\langle\left[\partial_\mu\delta A_\nu(x)
   -\partial_\nu\delta A_\mu(x)\right]^2\right\rangle.
\label{eq:(4.1)}
\end{equation}
The contraction of this by the propagator~\eqref{eq:(3.41)}
(see~Fig.~\ref{fig:1}) gives the leading large-$N$ result as\footnote{The
integration over~$A_{y0}$ in~Eq.~\eqref{eq:(3.24)} is implicitly assumed in
this expression. However, since this expression reduces
to~Eq.~\eqref{eq:(4.4)} in the large-$N$ limit~\eqref{eq:(1.1)} that is
independent of~$A_{y0}$, the integration over~$A_{y0}$ in~Eq.~\eqref{eq:(3.24)}
is trivial.}
\begin{align}
   \left\langle F_{\mu\nu}(x)F_{\mu\nu}(x)\right\rangle
   &=\frac{4\pi}{N}\int\frac{dp_x}{2\pi}\frac{1}{2\pi R}\sum_{p_y}\,
   \frac{2p^2\mathcal{L}(p)}{(p^2+4\Lambda^2)\mathcal{L}(p)^2
   +4(1-p_y^2/p^2)\mathcal{K}(p)^2}.
\notag\\
   &=\frac{4\pi}{N}\sum_{n=-\infty}^\infty\int\frac{d^2p}{(2\pi)^2}\,
   e^{ip_y2\pi Rn}
   \frac{2p^2\mathcal{L}(p)}{(p^2+4\Lambda^2)\mathcal{L}(p)^2
   +4(1-p_y^2/p^2)\mathcal{K}(p)^2},
\label{eq:(4.2)}
\end{align}
where in the second equality we have used Eq.~\eqref{eq:(3.7)}.

First, in the large-$N$ limit~\eqref{eq:(1.1)}, as shown
in~Appendix~\ref{sec:B}, we can set
\begin{equation}
   \Hat{\mathcal{L}}(p)\to0,\qquad
   \mathcal{K}(p)\to0.
\label{eq:(4.3)}
\end{equation}
Equation~\eqref{eq:(4.2)} thus reduces to (recall Eq.~\eqref{eq:(3.30)})
\begin{equation}
   \left\langle F_{\mu\nu}(x)F_{\mu\nu}(x)\right\rangle
   =\frac{4\pi}{N}\sum_{n=-\infty}^\infty\int\frac{d^2p}{(2\pi)^2}\,
   e^{ip_y2\pi Rn}
   \frac{2p^2}{(p^2+4\Lambda^2)\mathcal{L}_\infty(p)}.
\label{eq:(4.4)}
\end{equation}

Next, we extract the perturbative part from this expression. If we expand
$\mathcal{L}_\infty(p)$~\eqref{eq:(3.31)} and Eq.~\eqref{eq:(4.4)} with respect
to~$\Lambda^2/p^2$, the terms in positive powers of~$\Lambda^2$
[$\sim(\Lambda^2/p^2)^k$] are regarded as the non-perturbative part, because
$\Lambda^2\sim e^{-4\pi/\lambda_R}$. This reasoning tells us that the gluon
condensate in perturbation theory~(PT) is given by\footnote{The precise form of
this expression, which will be studied below, is
\begin{equation}
   \left\langle F_{\mu\nu}(x)F_{\mu\nu}(x)\right\rangle_{\text{PT}}
   =\frac{4\pi}{N}\sum_{n=-\infty}^\infty
   I_{M,n}
\label{eq:(4.5)}
\end{equation}
with $M\to\infty$, where
\begin{align}
   I_{M,0}&\equiv\sum_{k=0}^{M-1}\int_{|p|\leq q}\frac{d^2p}{(2\pi)^2}\,
   p^2\left[-\ln(p^2/q^2)\right]^k
   \left[\frac{\lambda_R(q)}{4\pi}\right]^{k+1},
\notag\\
   I_{M,n\neq0}&\equiv\sum_{k=0}^{M-1}\int_{|p|\leq q}\frac{d^2p}{(2\pi)^2}\,
   e^{ip_y2\pi Rn}p^2\left[-\ln(p^2R^2)\right]^k
   \left[\frac{\lambda_R(1/R)}{4\pi}\right]^{k+1},
\label{eq:(4.6)}
\end{align}
are given by the $M$th-order expansion of the function $1/\ln(p^2/\Lambda^2)$
with respect to~$\lambda_R(q)$ or~$\lambda_R(1/R)$ (see Eqs.~\eqref{eq:(4.12)}
and~\eqref{eq:(4.17)}). For the original expression for~$N\to\infty$,
Eq.~\eqref{eq:(4.4)},
\begin{equation}
   \left\langle F_{\mu\nu}(x)F_{\mu\nu}(x)\right\rangle_{\text{PT}}
   =\frac{4\pi}{N}\sum_{n=-\infty}^\infty I_n,
\label{eq:(4.7)}
\end{equation}
where
\begin{equation}
   I_n\equiv\int_{|p|\leq q}\frac{d^2p}{(2\pi)^2}\,
   e^{ip_y2\pi Rn}\frac{2p^2}{(p^2+4\Lambda^2)\mathcal{L}_\infty(p)},
\label{eq:(4.8)}
\end{equation}
one can rigorously prove that
\begin{equation}
   |I_n-I_{M,n}|=O(\lambda_R^{M+1}).
\label{eq:(4.9)}
\end{equation}
This shows that Eq.~\eqref{eq:(4.5)} with~$M\to\infty$ actually gives the
asymptotic expansion of~Eq.~\eqref{eq:(4.4)}.}
\begin{equation}
   \left\langle F_{\mu\nu}(x)F_{\mu\nu}(x)\right\rangle_{\text{PT}}
   =\frac{4\pi}{N}\sum_{n=-\infty}^\infty\int\frac{d^2p}{(2\pi)^2}\,
   e^{ip_y2\pi Rn}\left.
   \frac{p^2}{\ln(p^2/\Lambda^2)}\right|_{\text{expansion in $\lambda_R$}},
\label{eq:(4.10)}
\end{equation}
where we explicitly indicate that the integrand should be expanded
in~$\lambda_R$ in the perturbative evaluation. In this expression, we analyze
the $n=0$ term and the $n\neq0$ terms separately.\footnote{The large-$N$
expression~\eqref{eq:(4.4)} is convergent in the IR region and is real once the
UV divergence is regularized by a momentum cutoff; in this sense, the gluon
condensate in the large-$N$ expansion is an unambiguous object. Thus, the
ambiguity found in the following argument stems from the \emph{artifact\/} of
the perturbative evaluation. This means that the resurgence structure is
already assured in this quantity.}

\subsection{The $n=0$ term}
The $n=0$ term exhibits the quartic UV divergence and we thus introduce the UV
cutoff~$q$:
\begin{equation}
   \left\langle F_{\mu\nu}(x)F_{\mu\nu}(x)\right\rangle_{\text{PT, $n=0$}}
   =\frac{4\pi}{N}\int_{|p|\leq q}\frac{d^2p}{(2\pi)^2}\,
   \left.\frac{p^2}{\ln(p^2/\Lambda^2)}\right|_{\text{expansion in $\lambda_R$}}.
\label{eq:(4.11)}
\end{equation}
Then, noting
\begin{equation}
   \ln(p^2/\Lambda^2)=\ln(p^2/q^2)+\frac{4\pi}{\lambda_R(q)},
\label{eq:(4.12)}
\end{equation}
where $\lambda_R(q)$ is the renormalized coupling at the cutoff scale~$q$
(see~Eq.~\eqref{eq:(3.14)}), we have the perturbative expansion with respect
to~$\lambda_R(q)$:
\begin{equation}
   \left\langle F_{\mu\nu}(x)F_{\mu\nu}(x)\right\rangle_{\text{PT, $n=0$}}
   =\frac{4\pi}{N}q^4\sum_{k=0}^\infty\int_{|p|\leq1}\frac{d^2p}{(2\pi)^2}\,
   p^2(-\ln p^2)^k\left[\frac{\lambda_R(q)}{4\pi}\right]^{k+1}.
\label{eq:(4.13)}
\end{equation}
From this perturbative series, we define the corresponding Borel transform
as~Eq.~\eqref{eq:(1.4)},
\begin{align}
   B_{n=0}(u)&\equiv
   \frac{4\pi}{N}q^4\int_{|p|\leq1}\frac{d^2p}{(2\pi)^2}\,
   p^2\sum_{k=0}^\infty\frac{(-\ln p^2)^k}{k!}u^k
\notag\\
   &=\frac{1}{N}q^4\frac{-1}{u-2}.
\label{eq:(4.14)}
\end{align}
Thus, the Borel transform of the $n=0$ term of~Eq.~\eqref{eq:(4.10)} develops a
pole singularity at~$u=2$. We had to know this, because the $n=0$ term
in~Eq.~\eqref{eq:(4.10)} is basically identical to the loop integral appearing
in the scalar condensate in the 2D $O(N)$ non-linear sigma model in the
uncompactified space~$\mathbb{R}^2$, which suffers from the $u=2$ renormalon
ambiguity~\cite{David:1982qv}. See also~Ref.~\cite{Beneke:1998ui}.

Through the Borel integral~\eqref{eq:(1.5)}, the Borel singularity at~$u=2$
gives rise to the renormalon ambiguity on the gluon condensate (focusing only
on the $n=0$ term):
\begin{align}
   \left\langle F_{\mu\nu}(x)F_{\mu\nu}(x)
   \right\rangle_{\text{$n=0$, IR renormalon at $u=2$}}
   &\sim\int_0^\infty du\,B_{n=0}(u)\,e^{-4\pi u/\lambda_R(q)}
\notag\\
   &\sim\frac{1}{N}q^4e^{-8\pi/\lambda_R(q)}(\pm\pi i)
\notag\\
   &=\frac{1}{N}\Lambda^4(\pm\pi i),
\label{eq:(4.15)}
\end{align}
where the sign depends on how one avoids the pole singularity at~$u=2$ ($+$ for
a contour in the upper plane, $-$ for a contour in the lower plane). We note
that the ambiguity caused by the IR renormalon itself is \emph{independent\/}
of the UV cutoff~$q$ that we introduced; the last line of~Eq.~\eqref{eq:(4.15)}
does not refer to the scale~$q$.

\subsection{The $n\neq0$ terms}
The $n\neq0$ terms in~Eq.~\eqref{eq:(4.10)}, 
\begin{equation}
   \left\langle F_{\mu\nu}(x)F_{\mu\nu}(x)
   \right\rangle_{\text{PT, $n\neq0$}}
   =\frac{4\pi}{N}\sum_{n\neq0}\int\frac{d^2p}{(2\pi)^2}\,
   \left.e^{ip_y2\pi Rn}\frac{p^2}{\ln(p^2/\Lambda^2)}
   \right|_{\text{expansion in $\lambda_R$}},
\label{eq:(4.16)}
\end{equation}
are the Fourier transforms and thus UV convergent. This time, instead
of~Eq.~\eqref{eq:(4.12)}, we use
\begin{equation}
   \ln(p^2/\Lambda^2)=\ln(p^2R^2)+\frac{4\pi}{\lambda_R(1/R)},
\label{eq:(4.17)}
\end{equation}
where $\lambda_R(1/R)$ is the renormalized coupling at the scale~$1/R$. Then
the perturbative expansion with respect to~$\lambda_R(1/R)$ is given by
\begin{equation}
   \left\langle F_{\mu\nu}(x)F_{\mu\nu}(x)
   \right\rangle_{\text{PT, $n\neq0$}}
   =\frac{4\pi}{N}\frac{1}{R^4}\sum_{n\neq0}\sum_{k=0}^\infty
   \int\frac{d^2p}{(2\pi)^2}\,
   e^{ip_y2\pi n}p^2(-\ln p^2)^k
   \left[\frac{\lambda_R(1/R)}{4\pi}\right]^{k+1}.
\label{eq:(4.18)}
\end{equation}
The corresponding Borel transform is thus
\begin{align}
   B_{n\neq0}(u)&\equiv
   \frac{4\pi}{N}\frac{1}{R^4}\sum_{n\neq0}
   \int\frac{d^2p}{(2\pi)^2}\,
   e^{ip_y2\pi n}p^2\sum_{k=0}^\infty\frac{(-\ln p^2)^k}{k!}u^k
\notag\\
   &=\frac{1}{N}\frac{1}{R^4}2\pi^{2u-4}\zeta(4-2u)
   \frac{{\mit\Gamma}(2-u)}{{\mit\Gamma}(u-1)}.
\label{eq:(4.19)}
\end{align}
This Borel transform has a pole at~$u=2$,
\begin{equation}
   B_{n\neq0}(u)\stackrel{u\sim2}{\sim}
   \frac{1}{N}\frac{1}{R^4}\frac{1}{u-2},
\label{eq:(4.20)}
\end{equation}
and thus the corresponding renormalon ambiguity is given by
\begin{align}
   \left\langle F_{\mu\nu}(x)F_{\mu\nu}(x)
   \right\rangle_{\text{$n\neq0$, IR renormalon at $u=2$}}
   &\sim\int_0^\infty du\,B_{n\neq0}(u)\,e^{-4\pi u/\lambda_R(q)}
\notag\\
   &\sim\frac{1}{N}\frac{1}{R^4}e^{-8\pi/\lambda_R(1/R)}(\mp\pi i)
\notag\\
   &=\frac{1}{N}\Lambda^4(\mp\pi i).
\label{eq:(4.21)}
\end{align}
This is \emph{precisely opposite\/} to the ambiguity in~Eq.~\eqref{eq:(4.15)}
arising from the $n=0$ term. Thus, for the $u=2$ singularity, the contribution
from the compactification (i.e., terms present for a finite~$R$) cancels the
singularity that exists in the uncompactified space.

One can be skeptical about the above cancellation of the Borel singularities,
because our argument used different renormalization scales for the coupling
constant: $q$ for the $n=0$ term and $1/R$ for the $n\neq0$ terms. However,
this is just for simplicity of expressions. If we want, we may use a general
common mass scale~$\mu$ and make use of
\begin{align}
   \frac{\lambda_R(q)}{4\pi}
   &=\frac{\lambda_R(\mu)}{4\pi}
   \left[1-\ln(\mu^2/q^2)\frac{\lambda_R(\mu)}{4\pi}\right]^{-1},
\notag\\
   \frac{\lambda_R(1/R)}{4\pi}
   &=\frac{\lambda_R(\mu)}{4\pi}
   \left[1-\ln(\mu^2R^2)\frac{\lambda_R(\mu)}{4\pi}\right]^{-1},
\label{eq:(4.22)}
\end{align}
to obtain the perturbative series in $\lambda_R(\mu)$. One can show the
cancellation of the renormalon ambiguities at~$u=2$ with this renormalization
scale.\footnote{From Eq.~\eqref{eq:(1.5)}, we see that the change of the
coupling constant from~$\lambda_R(q)$ or~$\lambda_R(1/R)$ to~$\lambda_R(\mu)$
amounts to the change in the Borel transform $B(u)\to B(u)e^{Lu}$,
where $L=\ln(\mu^2/q^2)$ or~$\ln(\mu^2R^2)$. This change does not affect the
location of Borel singularities.}

\subsection{New Borel singularity at~$u=3/2$}
We showed that the Borel singularity at~$u=2$, which exists in the
uncompactified space, disappears. However, this is not the end of the story. We
note that the $\zeta$~function in~Eq.~\eqref{eq:(4.19)}, $\zeta(z)$, possesses
a simple pole at~$z=1$. This produces the Borel singularity at~$u=3/2$:
\begin{equation}
   B_{n\neq0}(u)\stackrel{u\sim3/2}{\sim}
   \frac{1}{N}\frac{1}{R^4}\left(-\frac{1}{\pi}\right)
   \frac{1}{u-3/2}.
\label{eq:(4.23)}
\end{equation}
Since $B_{n=0}(u)$~\eqref{eq:(4.14)} has no corresponding singularity, we
conclude that the perturbative part~\eqref{eq:(4.10)} possesses the renormalon
ambiguity at~$u=3/2$ as
\begin{equation}
   \left\langle F_{\mu\nu}(x)F_{\mu\nu}(x)
   \right\rangle_{\text{IR renormalon at $u=3/2$}}
   =\frac{1}{N}\frac{\Lambda^3}{R}\frac{1}{\pi}(\pm\pi i).
\label{eq:(4.24)}
\end{equation}
This is the net ambiguity of perturbative evaluation of the gluon condensate in
our system. Since this is proportional to~$1/R$, it is clear that this
renormalon ambiguity is peculiar to the compactified
space~$\mathbb{R}\times S^1$. Also, the location of the singularity~$u=3/2$ is
not dividable by the minimal bion action (corresponding to~$u=2$); we do not
know of any semi-classical interpretation of this renormalon ambiguity.

\subsection{Renormalon in a physical observable}
Since the gluon condensate exhibits the quartic UV divergence, the gluon
condensate itself may not be regarded as a physical observable. However, our
result implies that there is a physical observable whose perturbative
evaluation suffers from the renormalon ambiguity in the same way as the gluon
condensate. An explicit example is provided by the gradient
flow~\cite{Luscher:2010iy} and its small flow time
expansion~\cite{Luscher:2011bx}. For the ``$U(1)$ gauge field'' $A_\mu(x)$, we
introduce the gradient flow for~$t\geq0$ by
\begin{equation}
   \partial_t B_\mu(t,x)=\partial_\nu G_{\nu\mu}(t,x)
   +\alpha_0\partial_\mu\partial_\nu B_\nu(t,x),\qquad
   B_\mu(t=0,x)=A_\mu(x),
\label{eq:(4.25)}
\end{equation}
where $G_{\mu\nu}(t,x)=\partial_\mu B_\nu(t,x)-\partial_\nu B_\mu(t,x)$ is the
field strength of the flowed gauge field and~$\alpha_0$ is the ``gauge
parameter''~\cite{Luscher:2010iy}. Since this equation can be solved as
\begin{align}
   &B_\mu(t,x)
\notag\\
   &=A_{\mu0}
   +\int d^2x'\,\int\frac{dp_x}{2\pi}\frac{1}{2\pi R}\sum_{p_y}\,e^{ip(x-x')}
   \left[
   \left(\delta_{\mu\nu}-\frac{p_\mu p_\nu}{p^2}\right)e^{-tp^2}
   +\frac{p_\mu p_\nu}{p^2}e^{-\alpha_0tp^2}\right]
   \delta A_\nu(x'),
\label{eq:(4.26)}
\end{align}
the ``gluon condensate'' of the flowed gauge field is simply given by putting
the Gaussian factor~$e^{-2tp^2}$ to~Eq.~\eqref{eq:(4.2)}:
\begin{align}
   \left\langle G_{\mu\nu}(t,x)G_{\mu\nu}(t,x)\right\rangle
   &=\frac{4\pi}{N}\sum_{n=-\infty}^\infty\int\frac{d^2p}{(2\pi)^2}\,
   e^{ip_y2\pi Rn}
   \frac{2p^2\mathcal{L}(p)e^{-2tp^2}}{(p^2+4\Lambda^2)\mathcal{L}(p)^2
   +4(1-p_y^2/p^2)\mathcal{K}(p)^2}
\notag\\
   &\stackrel{N\to\infty,\text{PT}}{\to}
   \frac{4\pi}{N}\sum_{n=-\infty}^\infty\int\frac{d^2p}{(2\pi)^2}\,
   \left.e^{ip_y2\pi Rn}
   \frac{p^2e^{-2tp^2}}{\ln(p^2/\Lambda^2)}
   \right|_{\text{expansion in~$\lambda_R$}}.
\label{eq:(4.27)}
\end{align}
We then repeat the argument developed so far in this section. This time, the
$n=0$ term is UV convergent thanks to the Gaussian factor and we can take
$1/\sqrt{t}$ as the renormalization scale. Then the Borel sum gives
\begin{align}
   \left\langle G_{\mu\nu}(t,x)G_{\mu\nu}(t,x)\right\rangle_{\text{PT},n=0}
   &=\frac{1}{N}\frac{1}{t^2}\int_0^\infty du\,2^{u-2}{\mit\Gamma}(2-u)
   e^{-4\pi u/\lambda_R(1/\sqrt{t})}
\notag\\
   &\sim\frac{1}{N}\Lambda^4(\pm\pi i)+O(t),
\label{eq:(4.28)}
\end{align}
where in the second line, we indicate only the ambiguous part. In the $n\neq0$
terms, on the other hand, we may expand the Gaussian factor
as~$e^{-2tp^2}=1+O(t)$. The first term in this expansion is nothing but
Eq.~\eqref{eq:(4.18)}. Therefore, using Eq.~\eqref{eq:(4.19)},
\begin{align}
   \left\langle G_{\mu\nu}(t,x)G_{\mu\nu}(t,x)\right\rangle_{\text{PT},n\neq0}
   &=\frac{1}{N}\frac{1}{R^4}\int_0^\infty du\,2\pi^{2u-4}\zeta(4-2u)
   \frac{{\mit\Gamma}(2-u)}{{\mit\Gamma}(u-1)}
   e^{-4\pi u/\lambda_R(1/R)}+O(t)
\notag\\
   &\sim\frac{1}{N}\frac{\Lambda^3}{R}\frac{1}{\pi}(\pm\pi i)
   +\frac{1}{N}\Lambda^4(\mp\pi i)+O(\Lambda^6R^2)+O(t).
\label{eq:(4.29)}
\end{align}
In the sum of Eqs.~\eqref{eq:(4.28)} and~\eqref{eq:(4.29)}, the renormalon
ambiguity at~$u=2$ cancels out and the leading ambiguity (for~$\Lambda R\ll1$)
stems from the $u=3/2$ singularity.\footnote{In terms of the small flow time
expansion~\cite{Luscher:2011bx}, the $O(t^0)$ terms in Eqs.~\eqref{eq:(4.28)}
and~\eqref{eq:(4.29)} correspond to the gluon
condensate~$\langle F_{\mu\nu}(x)F_{\mu\nu}(x)\rangle$.} This example clearly
illustrates that the $u=3/2$ renormalon on~$\mathbb{R}\times S^1$ indeed
appears in the perturbative calculation of a physical observable.\footnote{It
might be possible to detect this leading renormalon on~$\mathbb{R}\times S^1$
by using the stochastic perturbation
theory~\cite{Bauer:2011ws,Bali:2014fea,DallaBrida:2016dai,DelDebbio:2018ftu,
Bruckmann:2019mky}.}

\section{Conclusion and discussion}
\label{sec:5}
In this paper, in the leading order of the large-$N$ approximation, we studied
the renormalon ambiguity in the gluon condensate in the 2D supersymmetric
$\mathbb{C}P^{N-1}$ model on~$\mathbb{R}\times S^1$ with the $\mathbb{Z}_N$
twisted boundary conditions. We found that the Borel singularity at~$u=2$,
which exists in the uncompactified space~$\mathbb{R}^2$, disappears in the
compactified space~$\mathbb{R}\times S^1$. Instead, we found an unfamiliar
singularity at~$u=3/2$, which is peculiar to the compactified space and has no
obvious semi-classical interpretation. We also showed that this renormalon
indeed appears in the perturbative calculation of a physical observable. We
emphasize that this result, which may be unexpected, was obtained by a very
straightforward and systematic calculation.

The $u=3/2$ singularity peculiar to the compactified space may be understood as
follows. For example, let us take Eq.~\eqref{eq:(4.27)} in the original form
with the discrete KK momentum:
\begin{equation}
   \left\langle G_{\mu\nu}(t,x)G_{\mu\nu}(t,x)\right\rangle_{\text{PT}}
   =\frac{4\pi}{N}\int\frac{dp_x}{2\pi}\frac{1}{2\pi R}\sum_{p_y}
   \left.\frac{p^2e^{-2tp^2}}{\ln(p^2/\Lambda^2)}
   \right|_{\text{expansion in~$\lambda_R$}}.
\label{eq:(5.1)}
\end{equation}
We note that in this form the integrand/summand possesses a singularity at an
IR point~$p^2=\Lambda^2$, which can be regarded as an indication of the IR
renormalon. For~$p_y\neq0$, however, we do not encounter this singularity
because for~$p_y=n/R$ with~$n\neq0$, $p^2\geq1/R^2\gg\Lambda^2$
for~$\Lambda R\ll1$. That is, the $S^1$ radius~$R$ acts as an IR cutoff
for~$p_y\neq0$. On the other hand, this singularity matters for~$p_y=0$. The
perturbative part in this $p_y=0$ sector reads
\begin{align}
   \left\langle G_{\mu\nu}(t,x)G_{\mu\nu}(t,x)\right\rangle_{\text{PT, $p_y=0$}}
   &=\frac{4\pi}{N}\int\frac{dp_x}{2\pi}\frac{1}{2\pi R}
   \left.\frac{p_x^2e^{-2tp_x^2}}{\ln(p_x^2/\Lambda^2)}
   \right|_{\text{expansion in~$\lambda_R$}}
\notag\\
   &=\frac{4\pi}{N}\frac{1}{2\pi R}\frac{1}{t^{3/2}}\sum_{k=0}^\infty
   \int\frac{dp_x}{2\pi}\,p_x^2e^{-2p_x^2}
   (-\ln p_x^2)^k\left[\frac{\lambda_R(1/\sqrt{t})}{4\pi}\right]^{k+1},
\label{eq:(5.2)}
\end{align}
and the corresponding Borel transform is
\begin{align}
   B_{p_y=0}(u)&=\frac{4\pi}{N}\frac{1}{2\pi R}\frac{1}{t^{3/2}}
   \int\frac{dp_x}{2\pi}\,p_x^2e^{-2p_x^2}
   \sum_{k=0}^\infty\frac{(-\ln p_x^2)^k}{k!}u^k
\notag\\
   &=\frac{1}{N}\frac{1}{Rt^{3/2}}
   \frac{1}{\pi}2^{u-3/2}{\mit\Gamma}(3/2-u).
\label{eq:(5.3)}
\end{align}
This precisely reproduces the singularity at~$u=3/2$ and the renormalon
ambiguity in~Eq.~\eqref{eq:(4.29)}, $(1/N)(\Lambda^3/R)(1/\pi)(\pm\pi i)$. The
number $u=3/2$ is thus naturally understood as the consequence of the
reduction of the momentum integration to one dimension. If the integration
measure in~Eq.~\eqref{eq:(5.3)} were $d^2p$, the leading singularity would be
$u=2$; the reduction of the spacetime dimension makes the IR divergence
stronger.

One might consider that the Borel singularity at~$u=3/2$ appears in a quantum
mechanical system that is obtained by the dimensional reduction of the present
system because this singularity stems from the lowest KK mode, i.e., the
$p_y=0$ sector. However, this singularity essentially originates from the
logarithmic factor in the integrand of, for instance, Eq.~\eqref{eq:(4.27)},
which comes from the running of the coupling constant. Such a running of the
coupling constant does not occur in quantum mechanics (at least naively) and
this phenomenon should be regarded as being peculiar to quantum field theory.
Thus, we do not expect this singularity in quantum mechanics.

We finally mention the relation between the present work and preceding analysis
based on the bion configuration. In~Ref.~\cite{Fujimori:2018kqp}, the vacuum
energy~$E$ is computed as a function of the supersymmetry breaking
parameter~$\delta\epsilon$ in
\begin{equation}
   \delta S
   \equiv\int d^2x\,\frac{\delta\epsilon}{\pi R}
   \sum_{a=1}^{N-1}
   m_a\left(\frac{|\varphi^a|^2}{1+\sum_{b=1}^{N-1}|\varphi^b|^2}
   -\frac{1}{N}\right)
   =\int d^2x\,\frac{\delta\epsilon}{\pi R}\sum_{A=1}^Nm_A
   \left(\Bar{z}^Az^A-\frac{1}{N}\right),
\label{eq:(5.4)}
\end{equation}
as
\begin{equation}
   E(\delta\epsilon)
   =E^{(0)}+E^{(1)}\delta\epsilon+E^{(2)}\delta\epsilon^2+\dotsb.
\label{eq:(5.5)}
\end{equation}
The leading ambiguity from the bion calculus was found in the~$E^{(2)}$ term.
Since these expansion coefficients can be obtained as the correlation
functions
\begin{align}
   E^{(1)}&=2\sum_Am_A\left\langle\Bar{z}^Az^A-\frac{1}{N}\right\rangle,
\notag\\
   E^{(2)}&=-\frac{1}{\pi R}\int d^2x\,
   \sum_Am_A\sum_Bm_B\left\langle
   \Bar{z}^Az^A(x)\Bar{z}^Bz^B(0)\right\rangle_c
\label{eq:(5.6)}
\end{align}
in the supersymmetric theory (where $E^{(0)}=0$), it is interesting to compute
these numbers with the large-$N$ technique developed in this paper. We hope to
come back to this problem in the near future.

\section*{Acknowledgements}
Discussions with
Toshiaki Fujimori,
Nobuyuki Ishibashi,
Tatsuhiro Misumi,
Norisuke Sakai,
and especially Kazuya Yonekura
at the YITP workshop ``Strings and Fields 2018'' motivated the present work.
This work was supported by JSPS Grant-in-Aid for Scientific Research Grant
Numbers JP18J20935 (O.M.), JP16H03982 (H.S.), and JP19K14711 (H.T.).

\appendix

\section{Basic formulas with the homogeneous coordinate variables}
\label{sec:A}
In this appendix, we summarize some useful formulas to
obtain~Eq.~\eqref{eq:(2.19)}. For the bosonic part of the action,
see~Ref.~\cite{Coleman:1985rnk}. In terms of the homogeneous coordinate, the
Fubini--Study metric~\eqref{eq:(2.2)} is written as
\begin{equation}
   G_{a\Bar{b}}=|z^N|^2(\delta^{ab}-\Bar{z}^az^b).
\label{eq:(A1)}
\end{equation}
Then, the fermionic fields satisfy the relation
\begin{equation}
   G_{a\Bar{b}}\Bar{\psi}_s^{\Bar{b}}\psi_{s'}^a=\Bar{\chi}_s^A\chi_{s'}^A,
\label{eq:(A2)}
\end{equation}
where~$s$, $s'=+$ or $-$. From the connection on~$\mathbb{C}P^{N-1}$,
\begin{equation}
   {\mit\Gamma}^a{}_{bc}
   =-\frac{\delta^{ab}\Bar{\varphi}^{\Bar{c}}+\delta^{ac}\Bar{\varphi}^{\Bar{b}}}
   {1+\sum_d|\varphi^d|^2},\qquad
   {\mit\Gamma}^{\Bar{a}}{}_{\Bar{b}\Bar{c}}
   =-\frac{\delta^{ab}\varphi^c+\delta^{ac}\varphi^b}{1+\sum_d|\varphi^d|^2},
\label{eq:(A3)}
\end{equation}
we have the kinetic term of the homogeneous coordinate variables~$\chi$
and~$\Bar{\chi}$ as
\begin{equation}
   G_{a\Bar{b}}\Bar{\psi}_\pm^{\Bar{b}}
   (\partial_\mu\psi_\pm^a+{\mit\Gamma}^a{}_{cd}\partial_\mu\varphi^c\psi_\pm^d)
   =\Bar{\chi}_\pm^A(\partial_\mu-ij_\mu)\chi_\pm^A.
\label{eq:(A4)}
\end{equation}
To translate the four-fermion interaction,
$R_{a\Bar{b}c\Bar{d}}\psi_+^a\Bar{\psi}_+^{\Bar{b}}\psi_-^c\Bar{\psi}_-^{\Bar{d}}$,
into that of the homogeneous coordinate, we note that the Riemann curvature
on~$\mathbb{C}P^{N-1}$ satisfies
\begin{equation}
   R_{a\Bar{b}c\Bar{d}}=-G_{a\Bar{b}}G_{c\Bar{d}}- G_{a\Bar{d}}G_{c\Bar{b}}.
\label{eq:(A5)}
\end{equation}
This relation and~Eq.~\eqref{eq:(A2)} immediately indicate
\begin{equation}
   R_{a\Bar{b}c\Bar{d}}\psi_+^a\Bar{\psi}_+^{\Bar{b}}\psi_-^c\Bar{\psi}_-^{\Bar{d}}
   =\chi_+^A\Bar{\chi}_-^A\chi_-^B\Bar{\chi}_+^B
   -\chi_+^A\Bar{\chi}_+^A\chi_-^B\Bar{\chi}_-^B.
\label{eq:(A6)}
\end{equation}

\section{Bounds for the functions $\Hat{V}(z)$, $\hat{\mathcal{L}}(p)$,
and~$\mathcal{K}(p)$}
\label{sec:B}
The modified Bessel function of the second kind $K_\nu(z)$ with~$\nu=0$ or~$1$
has the upper bound
\begin{equation}
   K_\nu(z)<\frac{2}{z}e^{-z/2},\qquad\text{for $z>0$}.
\label{eq:(B1)}
\end{equation}
This follows from the integral representation,
\begin{equation}
   K_\nu(z)=\frac{1}{2}\int_0^\infty dx\,
   x^{-\nu-1}e^{-\frac{z}{2}\left(x+\frac{1}{x}\right)},
\label{eq:(B2)}
\end{equation}
as
\begin{align}
   K_\nu(z)
   &=\frac{1}{2}\left(\int_0^1+\int_1^\infty\right)
   dx\,x^{-\nu-1}e^{-\frac{z}{2}\left(x+\frac{1}{x}\right)}
\notag\\
   &=\frac{1}{2}\int_0^1dx\,
   x^{-\nu-1}e^{-\frac{z}{2}\left(x+\frac{1}{x}\right)}
   +\frac{1}{2}\int_0^1dx\,
   x^{\nu-1}e^{-\frac{z}{2}\left(x+\frac{1}{x}\right)}
\notag\\
   &<\int_0^1dx\,x^{-2}e^{-\frac{z}{2}\frac{1}{x}}
   =\frac{2}{z}e^{-z/2},
\label{eq:(B3)}
\end{align}
where, in the second equality, we have changed the integration
variable~$x\to1/x$ in the second integral. The last inequality follows
from~$x^{\pm\nu-1}\leq x^{-2}$ for~$0\leq x\leq1$ and~$\nu=0$, $1$.
 
First, for~Eq.~\eqref{eq:(3.18)}, by using~Eq.~\eqref{eq:(B1)},
\begin{align}
   |\Hat{V}(z)|&\leq4\sum_{m\neq0}\frac{\sqrt{z}}{2\pi RN|m|}
   K_1(\sqrt{z}2\pi RN|m|)
\notag\\
   &<4\sum_{m\neq0}\frac{2}{(2\pi RN|m|)^2}
   e^{-\sqrt{z}\pi RN|m|}
\notag\\
   &<\frac{16}{(2\pi RN)^2}\frac{e^{-\sqrt{z}\pi RN}}{1-e^{-\sqrt{z}\pi RN}}.
\label{eq:(B4)}
\end{align}
Thus, $\Hat{V}(z)\to0$ under the large-$N$ limit~\eqref{eq:(1.1)}.

For~Eq.~\eqref{eq:(3.32)},
\begin{align}
   |\Hat{\mathcal{L}}(p)|
   &\leq\int_0^1dx\,
   \sum_{m\neq0}\frac{2\pi RN|m|}{\sqrt{\Lambda^2+x(1-x)p^2}}
   K_1(\sqrt{\Lambda^2+x(1-x)p^2}2\pi RN|m|)
\notag\\
   &<2\int_0^1dx\,
   \frac{1}{\Lambda^2+x(1-x)p^2}
   \sum_{m\neq0}e^{-\sqrt{\Lambda^2+x(1-x)p^2}\pi RN|m|}
\notag\\
   &=4\int_0^1dx\,
   \frac{1}{\Lambda^2+x(1-x)p^2}
   \frac{e^{-\sqrt{\Lambda^2+x(1-x)p^2}\pi RN}}
   {1-e^{-\sqrt{\Lambda^2+x(1-x)p^2}\pi RN}}
\notag\\
   &<\frac{4}{\Lambda^2}
   \frac{e^{-\Lambda\pi RN}}
   {1-e^{-\Lambda\pi RN}}.
\label{eq:(B5)}
\end{align}

For Eq.~\eqref{eq:(3.37)}, starting from
\begin{equation}
   |\mathcal{K}(p)|\leq
   \int_0^1dx\,
   \sum_{m\neq0}2\pi RN|m|
   K_0(\sqrt{\Lambda^2+x(1-x)p^2}2\pi RN|m|),
\label{eq:(B6)}
\end{equation}
a calculation parallel to the above leads to
\begin{equation}
   |\mathcal{K}(p)|
   <\frac{4}{\Lambda}
   \frac{e^{-\Lambda\pi RN}}
   {1-e^{-\Lambda\pi RN}}.
\label{eq:(B7)}
\end{equation}

The functions $\Hat{\mathcal{L}}(p)$ and $\mathcal{K}(p)$ thus vanish in the
large-$N$ limit~\eqref{eq:(1.1)}. (This is because the functions
$N^k|\Hat{\mathcal{L}}(p)|$ and~$N^k|\mathcal{K}(p)|$ are zero as~$N\to\infty$
for an arbitrary positive integer~$k$.)

\section{Emergence of $S_{\text{local}}$~\eqref{eq:(3.39)} with our convention of
$\gamma_5$ in dimensional regularization}
\label{sec:C}
In this appendix, we explain the essence for the emergence of the chiral
symmetry breaking term~\eqref{eq:(3.39)} with our convention of~$\gamma_5$
defined in footnote~\ref{footnote:4}. The expansion of the effective
action~\eqref{eq:(3.1)} with respect to the fluctuation~$\delta\sigma$
in~Eq.~\eqref{eq:(3.34)} to the quadratic order yields, just
as~Eq.~\eqref{eq:(3.27)},
\begin{align}
   &\left.S_{\text{eff}}\right|_{O(\delta\sigma^2)}
\notag\\
   &=-N\int\frac{dp_x}{2\pi}\,\frac{1}{2\pi R}\sum_{p_y}
   \left(-\frac{1}{2}\right)\sum_{m=-\infty}^\infty e^{-iA_{y0}2\pi RNm}
\notag\\
   &\qquad{}
   \times\int\frac{d^2k}{(2\pi)^2}\,e^{ik_y2\pi RNm}
   \tr\Bigl\{
   \frac{1}{i\Slash{k}+\Bar{\sigma}_0P_++\sigma_0P_-}
   \left[\widetilde{\delta\Bar{\sigma}}(p)P_+
   +\widetilde{\delta\sigma}(p)P_-\right]
\notag\\
   &\qquad\qquad\qquad\qquad\qquad\qquad\qquad{}
   \times
   \frac{1}{i(\Slash{k}-\Slash{p})+\Bar{\sigma}_0P_++\sigma_0P_-}
   \left[\widetilde{\delta\Bar{\sigma}}(-p)P_+
   +\widetilde{\delta\sigma}(-p)P_-\right]
   \Bigr\}.
\label{eq:(C1)}
\end{align}
For the $m=0$ term, which is potentially UV divergent, we apply dimensional
regularization setting $2\to D=2-2\varepsilon$. We adopt the convention
for~$\gamma_5$ defined in footnote~\ref{footnote:4}. Then we note
\begin{equation}
   \frac{1}{i\Slash{k}+\Bar{\sigma}_0P_++\sigma_0P_-}
   =(-i\Slash{k}+\Bar{\sigma}_0P_-+\sigma_0P_+)
   \frac{1}{k^2+i\Slash{\Hat{k}}(\sigma_0-\Bar{\sigma}_0)\gamma_5
   +\Bar{\sigma}_0\sigma_0},
\label{eq:(C2)}
\end{equation}
where $\Slash{\Hat{k}}\equiv\gamma_\mu\Hat{k}_\mu$ and $\Hat{k}_\mu$ takes
non-zero values only when~$\mu\neq x$ and~$\mu\neq y$ (i.e., $\Slash{\Hat{k}}$
\emph{commutes\/} with~$\gamma_5$). This
factor~$\Slash{\Hat{k}}(\sigma_0-\Bar{\sigma}_0)$ can contribute in the
limit~$\varepsilon\to0$ only when it is multiplied by the UV divergence
represented by~$1/\varepsilon$. However, the expansion of~Eq.~\eqref{eq:(C2)}
with respect to~$\Slash{\Hat{k}}(\sigma_0-\Bar{\sigma}_0)$ decreases the degree
of divergence and makes the integral in~Eq.~\eqref{eq:(C1)} finite. Thus, the
$m=0$ term in~Eq.~\eqref{eq:(C1)} can be written as
\begin{align}
   &\left.S_{\text{eff}}\right|_{O(\delta\sigma^2),m=0}
\notag\\
   &=-N\int\frac{dp_x}{2\pi}\,\frac{1}{2\pi R}\sum_{p_y}
   \left(-\frac{1}{2}\right)
\notag\\
   &\qquad{}
   \times\int_0^1dx\,\int\frac{d^Dk}{(2\pi)^D}\,
   \frac{1}{(k^2-2xkp+\Lambda^2+xp^2)^2}
\notag\\
   &\qquad\qquad{}
   \times
   \tr\Bigl\{
   (-i\Slash{k}+\Bar{\sigma}_0P_-+\sigma_0P_+)
   \left[\widetilde{\delta\Bar{\sigma}}(p)P_+
   +\widetilde{\delta\sigma}(p)P_-\right]
\notag\\
   &\qquad\qquad\qquad\qquad{}
   \times
   \left[-i(\Slash{k}-\Slash{p})+\Bar{\sigma}_0P_-+\sigma_0P_+\right]
   \left[\widetilde{\delta\Bar{\sigma}}(-p)P_+
   +\widetilde{\delta\sigma}(-p)P_-\right]
   \Bigr\}
\notag\\
   &=-N\int\frac{dp_x}{2\pi}\,\frac{1}{2\pi R}\sum_{p_y}
   \left(-\frac{1}{2}\right)
\notag\\
   &\qquad{}
   \times\int_0^1dx\,\int\frac{d^Dk}{(2\pi)^D}\,
   \frac{1}{(k^2-2xkp+\Lambda^2+xp^2)^2}
\notag\\
   &\qquad\qquad{}
   \times
   \tr\Bigl\{
   (-i\Slash{k}+\Bar{\sigma}_0P_-+\sigma_0P_+)
   (-i\Slash{\Hat{k}})
   \left[
   \widetilde{\delta\Bar{\sigma}}(p)\widetilde{\delta\Bar{\sigma}}(-p)P_+
   +\widetilde{\delta\sigma}(p)\widetilde{\delta\sigma}(-p)P_-
   \right]
   \Bigr\}
\notag\\
   &\qquad\qquad\qquad{}
   +(\text{normal terms}).
\label{eq:(C3)}
\end{align}
In the last expression, we have kept only the terms that resulted from the
property that $\Slash{\Hat{k}}$ commutes with~$\gamma_5$. These terms survive
in the $\varepsilon\to0$ limit and break the chiral symmetry. In fact, since
the trace over Dirac indices requires an even number of gamma matrices, we have
\begin{align}
   \left.S_{\text{eff}}\right|_{O(\delta\sigma^2),m=0}
   &=-N\int\frac{dp_x}{2\pi}\,\frac{1}{2\pi R}\sum_{p_y}
   \frac{1}{2}
   \left[
   \widetilde{\delta\sigma}(p)\widetilde{\delta\sigma}(-p)
   +\widetilde{\delta\Bar{\sigma}}(p)\widetilde{\delta\Bar{\sigma}}(-p)
   \right]
\notag\\
   &\qquad{}
   \times\int_0^1dx\,\int\frac{d^Dk}{(2\pi)^D}\,
   \frac{\Hat{k}^2}{(k^2-2xkp+\Lambda^2+xp^2)^2}
   +(\text{normal terms}),
\label{eq:(C4)}
\end{align}
and since
\begin{equation}
   \int\frac{d^Dk}{(2\pi)^D}\,
   \frac{k_\mu k_\nu}{(k^2-2xkp+\Lambda^2+xp^2)^2}
   =\frac{1}{4\pi}\frac{1}{2}\delta_{\mu\nu}\frac{1}{\varepsilon}
   +O(\varepsilon^0)
\end{equation}
and $\Hat{\delta}_{\mu\nu}\delta_{\mu\nu}=D-2=-2\varepsilon$ (here,
$\Hat{\delta}_{\mu\nu}$ takes non-zero values only when $\mu=\nu\neq x$ and
$\mu=\nu\neq y$; note that $\Hat{k}^2=\Hat{\delta}_{\mu\nu}k_\mu k_\nu$),
Eq.~\eqref{eq:(C4)} gives rise to the first line of~Eq.~\eqref{eq:(3.39)}
as~$\varepsilon\to0$. The second line of~Eq.~\eqref{eq:(3.39)} that is
chirally symmetric arises from the above normal terms and the $m\neq0$ terms
of~Eq.~\eqref{eq:(C1)}; it is included in~$S_{\text{local}}$ because we
summarized $S_{\text{eff}}|_{\text{quadratic}}$~\eqref{eq:(3.36)} in terms
of~$\widetilde{\delta R}(p)$ and~$\widetilde{\delta I}(p)$
in~Eq.~\eqref{eq:(3.35)}. This explains the emergence of the chiral symmetry
breaking term~$S_{\text{local}}$~\eqref{eq:(3.39)}. In any case, since we can
remove $S_{\text{local}}$ by a local counterterm as we did in the main text, this
artifact of the UV regularization has no physical significance.


\begin{thebibliography}{00}

\bibitem{Argyres:2012vv} 
  P.~Argyres and M.~\"Unsal,
  Phys.\ Rev.\ Lett.\  {\bf 109}, 121601 (2012)
  doi:10.1103/PhysRevLett.109.121601
  [arXiv:1204.1661 [hep-th]].

\bibitem{Argyres:2012ka} 
  P.~C.~Argyres and M.~\"Unsal,
  JHEP {\bf 1208}, 063 (2012)
  doi:10.1007/JHEP08(2012)063
  [arXiv:1206.1890 [hep-th]].

\bibitem{Dunne:2012ae} 
  G.~V.~Dunne and M.~\"Unsal,
  JHEP {\bf 1211}, 170 (2012)
  doi:10.1007/JHEP11(2012)170
  [arXiv:1210.2423 [hep-th]].

\bibitem{Dunne:2012zk} 
  G.~V.~Dunne and M.~\"Unsal,
  Phys.\ Rev.\ D {\bf 87}, 025015 (2013)
  doi:10.1103/PhysRevD.87.025015
  [arXiv:1210.3646 [hep-th]].

\bibitem{tHooft:1977xjm} 
  G.~'t Hooft,
  ``Can We Make Sense Out of Quantum Chromodynamics?,''
  Subnucl.\ Ser.\  {\bf 15}, 943 (1979).

\bibitem{Beneke:1998ui} 
  M.~Beneke,
  Phys.\ Rept.\  {\bf 317}, 1 (1999)
  doi:10.1016/S0370-1573(98)00130-6
  [hep-ph/9807443].

\bibitem{Eto:2004rz} 
  M.~Eto, Y.~Isozumi, M.~Nitta, K.~Ohashi and N.~Sakai,
  Phys.\ Rev.\ D {\bf 72}, 025011 (2005)
  doi:10.1103/PhysRevD.72.025011
  [hep-th/0412048].

\bibitem{Eto:2006mz} 
  M.~Eto, T.~Fujimori, Y.~Isozumi, M.~Nitta, K.~Ohashi, K.~Ohta and N.~Sakai,
  Phys.\ Rev.\ D {\bf 73}, 085008 (2006)
  doi:10.1103/PhysRevD.73.085008
  [hep-th/0601181].

\bibitem{Eto:2006pg} 
  M.~Eto, Y.~Isozumi, M.~Nitta, K.~Ohashi and N.~Sakai,
  J.\ Phys.\ A {\bf 39}, R315 (2006)
  doi:10.1088/0305-4470/39/26/R01
  [hep-th/0602170].

\bibitem{Bruckmann:2007zh} 
  F.~Bruckmann,
  Phys.\ Rev.\ Lett.\  {\bf 100}, 051602 (2008)
  doi:10.1103/PhysRevLett.100.051602
  [arXiv:0707.0775 [hep-th]].

\bibitem{Brendel:2009mp} 
  W.~Brendel, F.~Bruckmann, L.~Janssen, A.~Wipf and C.~Wozar,
  Phys.\ Lett.\ B {\bf 676}, 116 (2009)
  doi:10.1016/j.physletb.2009.04.055
  [arXiv:0902.2328 [hep-th]].

\bibitem{Bruckmann:2018rra} 
  F.~Bruckmann and S.~Lochner,
  Phys.\ Rev.\ D {\bf 98}, no. 6, 065005 (2018)
  doi:10.1103/PhysRevD.98.065005
  [arXiv:1805.11313 [hep-th]].

\bibitem{Brezin:1976wa} 
  E.~Br\'ezin, J.-C.~Le Guillou and J.~Zinn-Justin,
  Phys.\ Rev.\ D {\bf 15}, 1558 (1977).
  doi:10.1103/PhysRevD.15.1558

\bibitem{Lipatov:1976ny} 
  L.~N.~Lipatov,
  Sov.\ Phys.\ JETP {\bf 45}, 216 (1977)
  [Zh.\ Eksp.\ Teor.\ Fiz.\  {\bf 72}, 411 (1977)].

\bibitem{LeGuillou:1990nq} 
  J.~C.~Le Guillou and J.~Zinn-Justin,
  ``Large order behavior of perturbation theory,''
  Amsterdam, Netherlands: North-Holland (1990) 580 p. (Current physics - sources and comments

\bibitem{Bogomolny:1980ur} 
  E.~B.~Bogomolny,
  Phys.\ Lett.\  {\bf 91B}, 431 (1980).
  doi:10.1016/0370-2693(80)91014-X

\bibitem{ZinnJustin:1981dx} 
  J.~Zinn-Justin,
  Nucl.\ Phys.\ B {\bf 192}, 125 (1981).
  doi:10.1016/0550-3213(81)90197-8

\bibitem{Dunne:2015eaa} 
  G.~V.~Dunne and M.~\"Unsal,
  PoS LATTICE {\bf 2015}, 010 (2016)
  doi:10.22323/1.251.0010
  [arXiv:1511.05977 [hep-lat]].

\bibitem{Hongo:2018rpy} 
  M.~Hongo, T.~Misumi and Y.~Tanizaki,
  JHEP {\bf 1902}, 070 (2019)
  doi:10.1007/JHEP02(2019)070
  [arXiv:1812.02259 [hep-th]].

\bibitem{Fujimori:2018kqp} 
  T.~Fujimori, S.~Kamata, T.~Misumi, M.~Nitta and N.~Sakai,
  JHEP {\bf 1902}, 190 (2019)
  doi:10.1007/JHEP02(2019)190
  [arXiv:1810.03768 [hep-th]].

\bibitem{Witten:2010cx} 
  E.~Witten,
  AMS/IP Stud.\ Adv.\ Math.\  {\bf 50}, 347 (2011)
  [arXiv:1001.2933 [hep-th]].

\bibitem{Cristoforetti:2012su} 
  M.~Cristoforetti {\it et al.} [AuroraScience Collaboration],
  Phys.\ Rev.\ D {\bf 86}, 074506 (2012)
  doi:10.1103/PhysRevD.86.074506
  [arXiv:1205.3996 [hep-lat]].

\bibitem{Anber:2014sda} 
  M.~M.~Anber and T.~Sulejmanpasic,
  JHEP {\bf 1501}, 139 (2015)
  doi:10.1007/JHEP01(2015)139
  [arXiv:1410.0121 [hep-th]].

\bibitem{Fujimori:2016ljw} 
  T.~Fujimori, S.~Kamata, T.~Misumi, M.~Nitta and N.~Sakai,
  Phys.\ Rev.\ D {\bf 94}, no. 10, 105002 (2016)
  doi:10.1103/PhysRevD.94.105002
  [arXiv:1607.04205 [hep-th]].

\bibitem{Fujimori:2017oab} 
  T.~Fujimori, S.~Kamata, T.~Misumi, M.~Nitta and N.~Sakai,
  Phys.\ Rev.\ D {\bf 95}, no. 10, 105001 (2017)
  doi:10.1103/PhysRevD.95.105001
  [arXiv:1702.00589 [hep-th]].

\bibitem{Coleman:1985rnk} 
  S.~Coleman,
  ``Aspects of Symmetry : Selected Erice Lectures,''
  doi:10.1017/CBO9780511565045

\bibitem{Brezin:1977sv} 
  E.~Brezin, C.~Itzykson, G.~Parisi and J.~B.~Zuber,
  Commun.\ Math.\ Phys.\  {\bf 59}, 35 (1978).
  doi:10.1007/BF01614153

\bibitem{DAdda:1978dle} 
  A.~D'Adda, P.~Di Vecchia and M.~L\"uscher,
  Nucl.\ Phys.\ B {\bf 152}, 125 (1979).
  doi:10.1016/0550-3213(79)90083-X

\bibitem{Wess:1992cp} 
  J.~Wess and J.~Bagger,
  ``Supersymmetry and supergravity,''

\bibitem{Cremmer:1978bh} 
  E.~Cremmer and J.~Scherk,
  Phys.\ Lett.\  {\bf 74B}, 341 (1978).
  doi:10.1016/0370-2693(78)90672-X

\bibitem{tHooft:1972tcz} 
  G.~'t Hooft and M.~J.~G.~Veltman,
  Nucl.\ Phys.\ B {\bf 44}, 189 (1972).
  doi:10.1016/0550-3213(72)90279-9

\bibitem{David:1982qv} 
  F.~David,
  Nucl.\ Phys.\ B {\bf 209}, 433 (1982).
  doi:10.1016/0550-3213(82)90266-8

\bibitem{Luscher:2010iy} 
  M.~L\"uscher,
  JHEP {\bf 1008}, 071 (2010)
  Erratum: [JHEP {\bf 1403}, 092 (2014)]
  doi:10.1007/JHEP08(2010)071, 10.1007/JHEP03(2014)092
  [arXiv:1006.4518 [hep-lat]].

\bibitem{Luscher:2011bx} 
  M.~L\"uscher and P.~Weisz,
  JHEP {\bf 1102}, 051 (2011)
  doi:10.1007/JHEP02(2011)051
  [arXiv:1101.0963 [hep-th]].

\bibitem{Bauer:2011ws} 
  C.~Bauer, G.~S.~Bali and A.~Pineda,
  Phys.\ Rev.\ Lett.\  {\bf 108}, 242002 (2012)
  doi:10.1103/PhysRevLett.108.242002
  [arXiv:1111.3946 [hep-ph]].

\bibitem{Bali:2014fea} 
  G.~S.~Bali, C.~Bauer and A.~Pineda,
  Phys.\ Rev.\ D {\bf 89}, 054505 (2014)
  doi:10.1103/PhysRevD.89.054505
  [arXiv:1401.7999 [hep-ph]].

\bibitem{DallaBrida:2016dai} 
  M.~Dalla Brida and M.~L\"uscher,
  PoS LATTICE {\bf 2016}, 332 (2016)
  doi:10.22323/1.256.0332
  [arXiv:1612.04955 [hep-lat]].

\bibitem{DelDebbio:2018ftu} 
  L.~Del Debbio, F.~Di Renzo and G.~Filaci,
  Eur.\ Phys.\ J.\ C {\bf 78}, no. 11, 974 (2018)
  doi:10.1140/epjc/s10052-018-6458-9
  [arXiv:1807.09518 [hep-lat]].

\bibitem{Bruckmann:2019mky} 
  F.~Bruckmann and M.~Puhr,
  arXiv:1906.09471 [hep-lat].

\end{thebibliography}
\end{document}